%% file: main.tex
\documentclass[journal]{IEEEtran}
\usepackage{textcomp}
\usepackage[noadjust]{cite}

\usepackage[T1]{fontenc} % optional enhanced font encoding
\usepackage{amsmath}
\usepackage[cmintegrals]{newtxmath}
\usepackage{bm}
\usepackage{url}

\usepackage[colorinlistoftodos]{todonotes}
\usepackage{pgfplots}
\pgfplotsset{compat=1.13}
\usepackage[siunitx]{circuitikz}
\usetikzlibrary{arrows,positioning,shapes,backgrounds,fit}  
\usepackage{graphicx}
\usepackage{subcaption}
\usetikzlibrary{arrows}
\usepgfplotslibrary{polar}
\usepackage[bottom]{footmisc}

\definecolor{mblue}{rgb}{0,0.45,0.74}
\definecolor{mred}{rgb}{0.85,0.33,0.10}
\definecolor{mgreen}{rgb}{0.47,0.67,0.19}
\definecolor{myellow}{rgb}{0.93,0.69,0.13}
\definecolor{mpurple}{rgb}{0.4940, 0.1840, 0.5560}

\pgfplotsset{every axis/.append style={
                    label style={font=\small},
                    tick label style={font=\small},  
                    legend style = {font = \small}
                    }}

\begin{document}
\title{Harmonic Radar with Adaptively Phase-Coherent Auxiliary Transmitters}

\author{Anastasia~Lavrenko,~\IEEEmembership{Member,~IEEE,}
        James~K.Cavers,~\IEEEmembership{Fellow,~IEEE,}
        Graeme~K.~Woodward,~\IEEEmembership{Senior Member,~IEEE}
      % <-this % stops a space
\thanks{Manuscript received April 22, 2021.This work was partially supported by the Royal Society of New Zealand via Catalyst:Seeding grant CSG-FRI1802 and by the Scion Postdoctoral Scholarship S07084.}
\thanks{A. Lavrenko was with Scion (New Zealand Forest Research Institute), Christchurch 8041, New Zealand and is currently with the University of Twente, Enschede 7500 AE, Netherlands (e-mail: a.lavrenko@utwente.nl).}% <-this % stops a space
\thanks{J. K. Cavers is with the Wireless Research Centre, University of Canterbury,
 Christchurch 8140, New Zealand, and also with the School of Engineering
Science, Simon Fraser University, Burnaby, BC K1S 5B6, Canada (e-mail:
cavers@sfu.ca)}
\thanks{G. K. Woodward is with the Wireless Research Centre, University of
Canterbury, Christchurch 8140, New Zealand (e-mail: graeme.woodward@
canterbury.ac.nz).}
}

\maketitle

\begin{abstract}
In harmonic radar (HR), the radio frequency transmitter illuminates a nonlinear target (the tag), causing the return signal to consist of harmonics at multiples of the transmitted carrier frequency.  Of them, the second harmonic is usually the strongest and the one to which the receiver is tuned.  This frequency difference distinguishes the tag reflection from environmental clutter, which remains at the uplink (transmitter to tag) frequency. However, the passive nature of HR tags severely limits the reflected power, and therefore the range of the downlink (tag to receiver) path. We propose to increase the range and/or signal to noise ratio (SNR) by novel restructuring at the physical and signal levels. For this, we accompany the original transmitter with auxiliary transmitters able to send simple tones that are synchronized to arrive at the tag in phase, and we design the receiver to detect an intermodulation component. The resulting range and SNR are much greater than those of the original, conventional HR system, and greater even than if the original system were to transmit with power equal to the aggregate power of our new system. Achieving mutually coherent, i.e., in phase, arrival of the tones at the tag is the focus of the present paper. We provide a system framework that models the tag and the uplink and downlink, then present the adaptive phase coherence algorithm and analyze the probabilistic growth of the output signal power. We also account for the effects of frequency shifts due to transmitter mobility and the frequency offset errors in the transmitter local oscillators. 
\end{abstract}


\IEEEpeerreviewmaketitle

\section{Introduction}
\input{sec_intro}

\section{Harmonic Radar}
\label{sec:2}
\input{sec_HR}

\subsection{Distributed HR System with Auxiliary (Helper) Nodes}
\label{sec:DS}
\input{sec_DS}

\section{Benefits of Using Helper Nodes}
\label{sec:TS}
\input{sec_TS}

\section{Phase Adaptation for Coherent Helper Node Signal Combining at the Tag}
\label{sec:PA}
\input{sec_PA}

\section{ Phase Adjustment Accuracy}
\input{sec_AN}

\section{Numerical Example}
\label{sec:Num}
\input{sec_Num}

\section{Conclusions}
The weak return signal in harmonic radar severely limits its range.  The conventional remedy is ever-higher power at the uplink transmitter, but the inverse sixth power dependence on range in HR makes this an expensive method.  In contrast, our approach is based on new physical and signal configurations.  Unconventionally, we use ``helper nodes'' to transmit tones that arrive as a single composite tone at the nonlinear tag while the ranging node transmits its own signal. 
Adding $M$ helper nodes increases the achievable range by a factor of $\sqrt[3]{2M}$,  giving a $100\%$  increase for 4 helper nodes and a $152\%$ range increase for 8 helpers. For this benefit, the ranging node (RN) receiver must correlate against transmitted signal $x(t)$, instead of  $x^2(t)$, but would not have to increase its transmit power.  In contrast, a conventional HR system would have to increase the transmit power of its RN in order to increase its range.  For a proper comparison, the conventional RN would be allotted the aggregate power of the helper-based system ($M$ helper nodes plus the RN). 
It would then experience a range extension factor of   $\sqrt[3]{M+1}$. Numerically, this is a range increase of only $71\%$ for $M=4$, and $108\%$ for $M=8$, significantly lower than that of the helper-based system.
This benefit however, requires the helper tones to arrive coherently, at the tag.  Ensuring that condition is the focus of this paper.

   Our phase-adaptation algorithm causes each helper, one at a time, to adjust its phase at the tag, aligning it with the composite tone from previously-aligned helpers.  
   Receiver noise causes adjustment error and consequent statistical variation in the composite helper sum at the tag.  Our analysis, which provides the closed form of the helper sum distribution for $M=2$ and a recursion for $M>2$, demonstrates a long-tailed distribution in which the 50th percentile is close to the maximum value, but the 10th percentile can be much lower, depending on the SNR and $M$. 
   Assessment of the method based on a measurement set from a prototype HR ranging node \cite{sto2020low} %.  
   showed that a single helper ($M=1$) provides $26\%$ more range, although a conventional system with doubled power can do the same.  Adding a second helper causes the 50th percentile range to increase more quickly than can the range of a conventional system with equal total power. However, the noise-affected 10th percentile range can actually shrink, because there is some probability that the two helper phases might oppose each other.  Additional helpers increase the 10th percentile range and, with 5 and more helpers, even the $10$th percentile range grows beyond the range of the conventional system. 
    An alternative to increasing the range is to reduce the uplink and downlink antenna gains.  For example, with $M = 4$ helper nodes and phase coherence (17), the SNR improvement at the Rx is $18$ dB.  Then the transmit and received antenna gains can each be reduced by 6 dB.  Since gains of the RN antennae in HR are typically at least 15 dBi, a reduction to 9 dBi can provide savings in size and cost.
   
    In the pursuit of range increase, it is an implementation issue whether our helper-based system or the conventional system is preferable. We note that conventional ranging nodes become bigger, heavier and costlier as their transmit power increases, and they may require additional linearization. In contrast, the helper-based system can add power incrementally by adding nodes as needed, without requiring upgrades to the RNs. Moreover, for systems with multiple RNs, a single set of helpers boosts the effective power of every RN.
   The success of the helper-node structure prompts questions that should be addressed in further research.  One issue is that powerful tones, as emitted from the helpers, are generally unwelcome near other electronics.  Are there alternative helper signals that have a lower power spectral density, do not interfere with the ranging application and still allow easy separation of intermodulation components for estimate of phase errors, or other property?  Another question is additional application areas.  Could our adaptive phase coherence at a point in space be used for, say, backscatter communications? 

%\newpage

%\appendix[]



\ifCLASSOPTIONcaptionsoff
  \newpage
\fi



\bibliographystyle{IEEEtran}
\bibliography{BibL_HR}

\clearpage

%\begin{IEEEbiographynophoto}{James K. Cavers}

%\newpage

%\begin{IEEEbiographynophoto}{Graeme K. Woodward}


%\vfill


\end{document}

%% file: sec_intro.tex
\IEEEPARstart{H}armonic radar (HR) is the basis of many schemes that interrogate, or track the location of, a simple nonlinear tag \cite{martone2016overview, gu2018diplexer, hui2019radio, kumar2019harmonic, nourshamsi2020joint}. Tracking of insect movement \cite{o2004tracking} and discovery of nonlinear electronic circuitry \cite{7811243} are typical applications. HR interrogation schemes are also used in a variety of IoT applications. Common examples include temperature and humidity sensing \cite{kubina2014quasi, lazaro2014passive}, vital sign monitoring \cite{singh2011respiratory, chioukh2014noise} and critical asset inspection \cite{palazzi2015demonstration, abdelnour2018passive}. 

The tag in HR has no power source of its own (i.e., it is passive) and, in its simplest form, it is just an antenna in series with a diode \cite{rasilainen2015transponder, lav2019design}. In the uplink, a node containing a radio transmitter (the Tx) sends a bandpass signal that illuminates the tag, inducing a voltage across the diode that in turn produces a multiband current according to the nonlinear diode equation. This results in a weak re-radiated signal at harmonics of the the incident wave frequency (now the fundamental frequency). A node containing a receiver (the Rx) detects this downlink return signal, typically processing only the second harmonic, which is usually the strongest. Because the downlink signal is so weak, the tag-Rx distance is limited, which often becomes the central problem in the system design. 
 In more detail, the second-harmonic downlink signal emitted by the tag has power proportional to the square of the power of the incident uplink signal, except at extremely close range. When combined with the typical inverse-square law dependence on distances in the uplink and downlink, it gives the overall signal power response from Tx to a co-located Rx an inverse sixth-power dependence on distance \cite{martone2016overview}. The result is a very short range. For instance, in \cite{sto2020low} a system using classical dipole-based tags was shown to have a detection range of up  to $40$m when operating in the S-band (2.9/5.8GHz) with 13dBi Tx/Rx antennas and an output power of 3W, and up to $15$m at $10$W with 15dBi antennas in the X-band (9.3/18.6GHz).
  
To increase the range, the usual approach is to increase the Tx power, since the diode's approximate squaring action results in a squared increase in power in the downlink.  Unfortunately, that range improvement is modest: when the power-squaring is combined with the inverse sixth-power distance dependence, the range varies as the cube root of any Tx power increase. For example, an $8$-fold increase in Tx power results in a corresponding range increase factor of just $\sqrt[3]{8} = 2$. Nevertheless, it is common to employ a very powerful transmitter or a high gain antenna at the Tx, or both, as a brute force way to increase the range. Examples of such high-power solutions include 25kW systems in \cite{riley2002design, milanesio2016design};  4kW and 3kW systems in \cite{colpitts2004harmonic} and \cite{tsai2012high}, respectively; and, most recently, a 1kW system in \cite{maggiora2019innovative}.

We take an alternative, and novel, approach to increasing the range of harmonic radar.  Recently, we proposed the use of auxiliary Tx nodes, which we term ``helper nodes'' because of their simplicity \cite{lav2019on}.   Although they transmit only simple tones, they provide a significant increase in average power of the desired signal at  Rx, and therefore allow a greater range or lower antenna gain, while being far simpler and less costly than the high-power Tx. The operation of a helper-based system is enabled by the nonlinear action of the tag that produces intermodulation terms when excited by multiple signal sources. Since the power of the intermodulation terms is defined by the product of the ranging signal and the helper node contribution, using it for ranging offers an opportunity for boosting the tag output. In a basic configuration, the helper tones are allowed to arrive at the tag incoherently.  Most of the time, this provides a significant increase in average power of the desired signal at the Rx.  During occasional phase misalignments, however, the received power can be very low. 
In this paper, we show how the helper tones can instead be made to arrive in phase, i.e., coherently, at the tag, thereby maximizing the power of the tag downlink and, in turn, maximizing the  Rx signal power and/or the range. 
In the following, we describe the adaptive phase coherence algorithm in detail, and provide an analysis of the resulting Markovian growth of signal power.  We analyze the performance improvement on an example of a practically built conventional system and explain the implications for the system designs. Our phase coherence algorithm may also find use in other applications where there is a need to phase-align multiple signal sources at a single point in space.

The general concept of a multi-tone harmonic radar that uses the intermodulation product at the output of a non-linear target has been previously explored in \cite{mazzaro2014detection, owen2018nonlinear, owen2019nonlinear}.
However, to the best of our knowledge there has been no prior work on the use of the intermodulation products of multiple helper transmitters to increase the downlink power from a passive, non-linear tag. Somewhat related is a design for backscatter data communication \cite{de2020analysis}, in which a desired data signal from a Tx arrives at a separate Rx, as does a simple tone from a third node (the exciter), resulting in beneficial interference at the Rx that puts the composite signal in a higher-gain region of the incoherent detector.
In contrast, our helper-node method performs ranging in harmonic radar at twice the carrier frequency of the uplink, the tag is simple and entirely passive, and the intermodulation component produced by helper tones is the essential part of the received signal. 

In what follows, Section II outlines the physical and signalling layout of the HR system. Section III then contrasts performance with phase-coherent helper nodes against brute force increase of Tx power.  Section IV defines the phase-coherence algorithm, and Section V provides the analyses of convergence, performance effects of the result, and limitations of the helper-node method.  Section VI then demonstrates the degree of improvement due to helper nodes for a specific implementation of a Tx-Rx ranging node. Finally, Section VII presents our conclusions about the helper-node method and what the next steps might be.  

%% file: sec_HR.tex
\subsection{Conventional Harmonic Radar}
\label{sec:sec2A}
\begin{figure}[t!]
         \centering
    \includegraphics[width=0.95\linewidth]{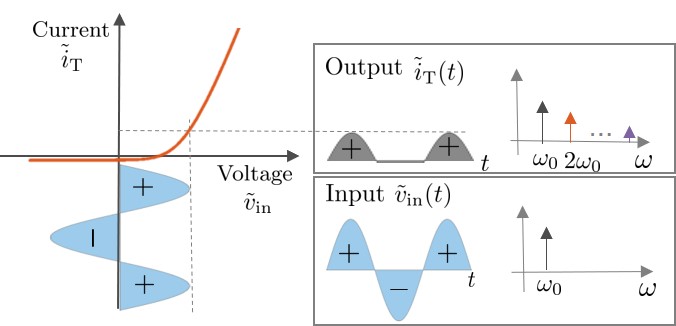}
  %   \caption{}
  %   \end{subfigure}
     \caption{\small{
     Relationship between the input voltage and the output current of a harmonic transponder tag, in time and frequency.}}
              \label{fig:Diode}
    \vspace{-0.2cm}
\end{figure}

In classical radar, the radar transmitter emits an RF pulse at frequency $\omega_{\rm 0} = 2\pi f_{\rm 0}$ and the radar receiver listens for reflections from a passive target at the same frequency. In contrast, a HR target is nonlinear, as well as passive, so its response is rich in harmonics.
Typically, such a harmonic response is induced by  attaching a battery-less harmonic transponder tag to the target of interest. HR tags typically combine  a resonant antenna, a low-voltage diode and possibly a simple impedance matching network \cite{colpitts2004harmonic, rasilainen2015transponder, lav2019design}.  The antenna voltage drives the diode, creating a rectified current that contains harmonics of the received signal, as sketched in Fig.~\ref{fig:Diode}.  The second harmonic is usually the strongest, so the tag's transmit antenna is  tuned to  $2\omega_0$. The second harmonic signal is thus emitted from the tag and subsequently detected at the HR receiver.
The main advantage of harmonic operation is that the background clutter is greatly reduced, since radio frequency reflection from most objects is linear, producing a backscattered response only  at $\omega_0$.  

Below, we provide a model of conventional HR that combines the effects of the uplink (HR transmitter to tag antenna input), the tag (tag antenna input, nonlinearity and tag antenna output), and the downlink (tag antenna output to the HR receiver). We assume for simplicity that the HR transmitter is collocated with the HR receiver, forming a single HR node (the ranging node, or RN).  Note that ranging systems can have more than one RN, e.g., in order to locate the tag by multilateration. For simplicity of presentation, our analysis considers just one.  Throughout, we also distinguish real bandpass signals from their complex envelopes by a tilde; for example, we denote the real bandpass signal transmitted at $\omega_0$ by $\tilde{s}(t) = \mathrm{Re}\{s(t) e^{\jmath \omega_0t}\}$ and  its baseband complex envelope by ${s}(t)$.
 
 In the uplink, the RN sends a pulse with complex envelope
 \begin{equation}
     s_{\rm r}(t) = \sqrt{2P_{\rm r}R_{\rm tx}}x(t),
     \label{eq:s_r}
 \end{equation}
 where $P_{\rm r}$ is the transmit RF power and $R_{\rm tx}$ is the transmit antenna resistance. Furthermore, we assume that the ranging signal $x(t)$ is a biphase
$(+1,-1)$ sequence with good autocorrelation properties.
  With line of sight transmission, the RN signal is received at the tag antenna with the complex envelope 
 \begin{equation}
     v_{\rm in}(t) = \sqrt{k_{\rm in} \frac{R_{\rm F}}{R_{\rm tx}}} h_{\rm u}(d_{\rm r}) s_{\rm r} (t-\tau_{\rm r})e^{\jmath \theta_{\rm r}},
     \label{eq:v_in}
 \end{equation}
 in which $R_{\rm F}$ and $k_{\rm in}$ are the effective tag antenna resistance and the input tag power transfer efficiency at $\omega_0$, respectively,  $h_{\rm u}(d_{\rm r})= \sqrt{G_{\rm tx} G_{\rm tag}(\omega_0) \left( {c}/{2\omega_0 d_{\rm r}}\right)^2}$ is the uplink gain with $G_{\rm tx}$, $ G_{\rm tag}(\omega)$ denoting the transmitter and tag antenna gains, respectively, $c$ is the speed of light, and $d_{\rm r}$ the distance from the RN to the tag. The propagation delay is $\tau_{\rm r} =  {d_{\rm r}}/{c}$ and the corresponding phase shift is $\theta_{\rm r} = -\omega_0 \tau_{\rm r}$.

 The tag can be modeled as a series circuit with the real bandpass voltage $\tilde{v}_{\rm in}(t) = \mathrm{Re}\{v_{\rm in}(t) e^{\jmath \omega_0t}\}$ as input and real multiband current $\tilde{i}_{\rm T}(t)$ as response, linked by the nonlinear equation
\begin{equation}
    \frac{\tilde{v}_{\rm in}(t)}{n_{\rm i}V_{\rm T}} = \rho \frac{\tilde{i}_{\rm T}(t)}{I_{\rm s}}+\ln \left( \frac{\tilde{i}_{\rm T}(t)}{I_{\rm s}}+1\right).
    \label{eq:diode_eq}
\end{equation}
Here, $n_{\rm i}$ and $V_{\rm T}$ are diode thermal voltage and  the ideality factor, respectively, $I_{\rm s}$ is its saturation current and $\rho = I_{\rm s}R_{\rm F}/n_{\rm i}V_{\rm T}$. An explicit solution to \eqref{eq:diode_eq} has been obtained in \cite{lav2020two-region} in terms of the Lambert W-function. From that, \cite{lav2020two-region} also provides the closed-form expressions for the relation between the complex envelope $v_{\rm in}(t)$ at the fundamental frequency $\omega_0$ and the complex envelope $i_2(t)$  of the current at the second harmonic $2\omega_0$, in two conditions:
\begin{itemize}
  \setlength\itemsep{0.5em}
    \item small-signal conditions ($v_{\rm in}(t)/n_iV_{\rm T} < -1- \ln \rho - \rho$), where
\begin{equation}
    i_{2}(t) \approx  \frac{\beta}{R_{\rm F}} v_{\rm in}^2(t),
      \label{eq:second_complx}
\end{equation}
in which
$\beta =  \frac{1}{4(n_iV_{\rm T})} \sum_{n=1}^{\infty} {n^{n
+1}} \frac{(-1)^{n-1}}{n!} \rho^n e^{n\rho}$. For $\rho < 0.04$ it simplifies further to $\beta \approx \rho/4n_{\rm i}V_{\rm T}$;

\vspace*{0.1cm}

      \item large-signal conditions ($v_{\rm in}(t)/n_iV_{\rm T} \gg 1$), where
\begin{equation}
    i_{2}(t) \approx \frac{2}{3\pi} \frac{1}{R_{\rm F}} |v_{\rm in}(t)| e^{\jmath 2\varphi_{\rm in}(t)},
    \label{eq:current_second_large}
\end{equation}
in which $\varphi_{\rm in}(t) = \arg(v_{\rm in}(t))$.
\end{itemize}
From \eqref{eq:second_complx} and \eqref{eq:current_second_large}, the phase of the tag output current at the second harmonic is double that of the input signal, while its magnitude grows quadratically in the small-signal region and linearly in large signal conditions.  
Our principal interest is to extend the limits of the operating range where signals at the tag are weak.  Consequently, in the following we consider the small-signal quadratic\footnote{However, our numerical results in Section~\ref{sec:num3} show that the phase-adaptation approach that we develop based on the small-signal model continues to work well in the tag regions above quadratic.} model \eqref{eq:second_complx}.

 With $R_{\rm H}$ as the effective output impedance of the tag at $2\omega_0$, the complex envelope of the tag output voltage in the small-signal conditions becomes
\begin{equation}
    v_{\rm out} = R_{\rm H} i_2(t) \approx \frac{R_{\rm H}}{R_{\rm F}} \beta v_{\rm in}^2(t),
    \label{eq:tag_output}
\end{equation}
with the corresponding bandpass equivalent provided by $\tilde{v}_{\rm out}(t) = (\beta R_{\rm H}/R_{\rm F}) \mathrm{Re}\{ v_{\rm in}^2(t) e^{\jmath 2\omega_0t}\}$. Note that for the rest of the paper, we drop the approximation sign in \eqref{eq:tag_output}, bearing in mind that it is a small-signal approximation. 

Finally, the downlink mirrors the uplink, such that at the HR receiver we obtain  
\begin{equation}
    r(t) = h_{\rm d}\sqrt{k_{\rm out}\frac{ R_{\rm rx}}{R_{\rm H}}} v_{\rm out}(t-\tau_{\rm r}) e^{\jmath 2\theta_{\rm r}} + n(t),
    \label{eq:received_sig}
\end{equation}
where $h_{\rm d} = \sqrt{G_{\rm rx} G_{\rm tag}(2\omega_0) \left( {c}/{4\omega_0 d_{\rm r}}\right)^2}$ is the downlink gain, in which  $G_{\rm rx}$  is the receiver antenna gain, $R_{\rm rx}$ is the resistance of the Rx antenna, and $k_{\rm out}$ is the output tag power transfer efficiency at $2\omega_0$.  The complex noise $n(t)$ in \eqref{eq:received_sig} is considered to be white Gaussian with power spectral density (PSD) $N_0 = R_{\rm rx}k_{\rm B}T_{\rm n}$ where $k_{\rm B}$ is the Boltzman constant and $T_{\rm n}$ is the noise temperature.

For notational convenience, we combine \eqref{eq:s_r} and \eqref{eq:v_in} to represent the input voltage at the tag as
\begin{equation}
    v_{\rm in}(t) = A_{\rm r}x(t-\tau_{\rm r})e^{\jmath \theta_{\rm r}},
    \label{eq:v_in2}
\end{equation}
where $A_{\rm r}=\sqrt{2R_{\rm F}k_{\rm in} P_{\rm r}} h_{\rm u}(d_{\rm r})$. Lastly, we also combine \eqref{eq:tag_output}, \eqref{eq:received_sig} and \eqref{eq:v_in2} so that
\begin{equation}
    r(t) = \eta  A_{\rm r}^2 x^2(t-2\tau_{\rm r})e^{\jmath 4\theta_{\rm r}} + n(t),
    \label{eq:recieved_sig2}
\end{equation}
where $\eta = h_{\rm d}(d_{\rm r})\beta \sqrt{R_{\rm H}R_{\rm rx}k_{\rm out}}/R_{\rm F}$. The product $\eta A_{\rm r}^2$ represents the total gain of the Tx--tag--Rx link.
In the absence of noise, \eqref{eq:v_in}, \eqref{eq:tag_output} and \eqref{eq:received_sig} combined make the received amplitude $|r(t)|$ proportional to $h_{\rm u}^2(d_{\rm r})h_{\rm d}(d_{\rm r})$, and hence inversely proportional to $d_{\rm r}^3$. The received power in harmonic radar is then inversely proportional to $d_{\rm r}^6$, i.e., an inverse sixth-power law.

%% file: sec_DS.tex
 \begin{figure}[t!]
% \vspace*{-\baselineskip}
         \centering
   \includegraphics[width = 1\linewidth]{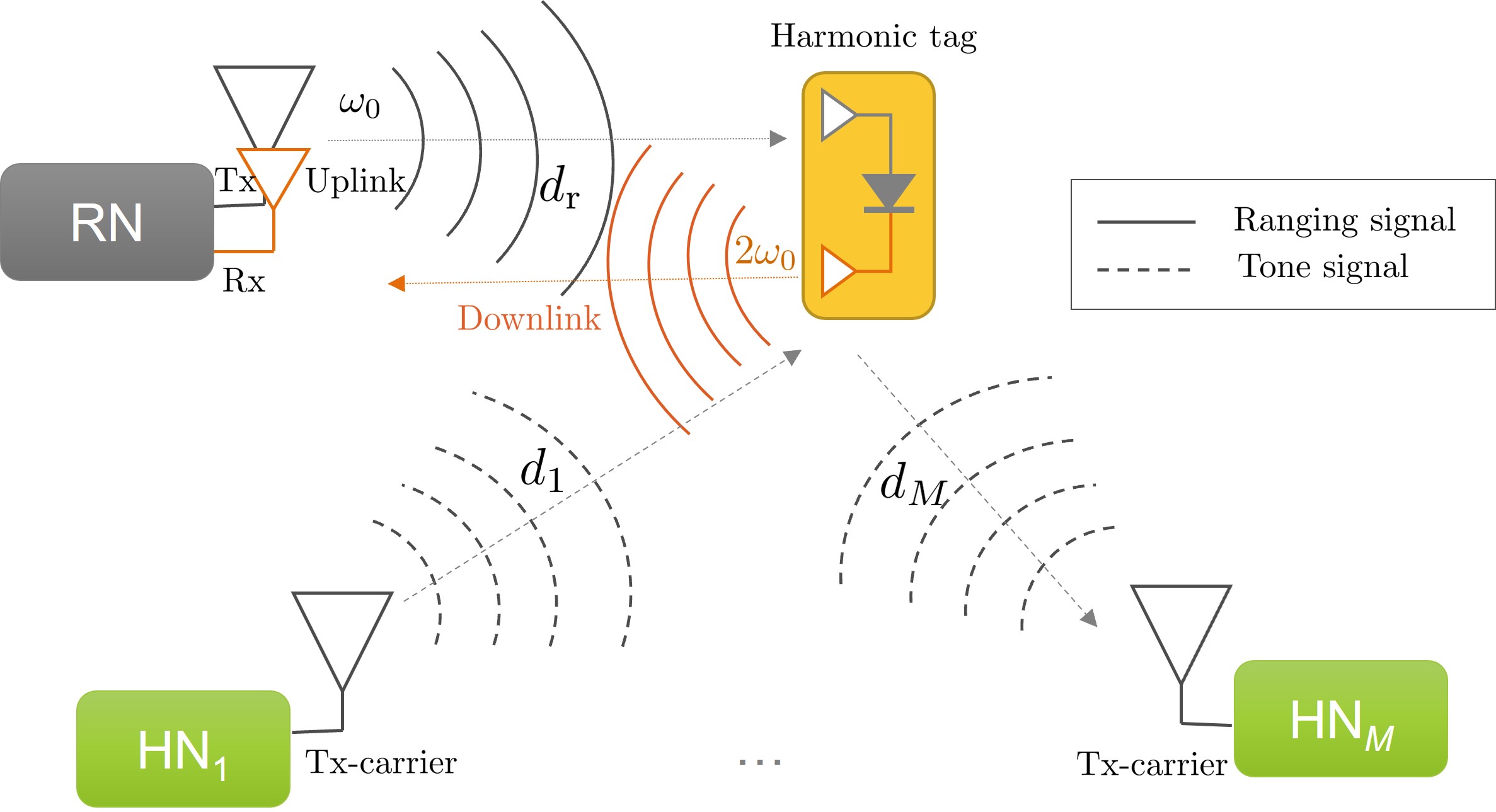}
     %   \vspace{0.3cm}
    \caption{\small{Harmonic radar system with auxiliary helper nodes: a single ranging node (RN) that transmits a ranging signal at $\omega_0$ and listens for signal returns at $2\omega_0$ aided by $M$ helper nodes (HNs) that transmit simple tones at $\omega_0$.}}
    \label{fig:HR_multist}
    \vspace{-0.2cm}
\end{figure}

Consider the system of \cite{lav2019on} that, in addition to the RN of a conventional HR system, employs $M$ auxiliary transmitters, each sending a simple tone at $\omega_0$ (see Fig.~\ref{fig:HR_multist}). For simplicity, we assume that these additional nodes, which we also refer to as helper nodes (HNs) or simply helpers, have the same transmit power and antennas as the ranging node.  Then, when the $m$-th helper sends the complex envelope $s_m = \sqrt{2P_{\rm r}R_{\rm tx}}e^{\jmath \theta_m}$, where $\theta_m$ denotes its local oscillator (LO) phase relative to that of the RN, the tag receives
\begin{align}
   v_{\rm in}(t) = \sqrt{k_{\rm in}\frac{R_{\rm F}}{R_{\rm tx}}} &h_{\rm u}(d_{\rm r}) s_{\rm r}(t-\tau_{\rm r})e^{\jmath \theta_{\rm r}} \notag \\
   & + \sqrt{2k_{\rm in}P_{\rm r}R_{\rm F}}\sum_{m=1}^Mh_{\rm u}(d_m)  e^{\jmath \bar{\theta}_m}.
    \label{eq:rtag}
\end{align}
Here, $d_m$ denotes the distance between the $m$-th helper node and the tag, while $\bar{\theta}_m = \theta_m-\omega_0\tau_m = \theta_m-\omega_0d_m/c$. With the notation of \eqref{eq:v_in2}, \eqref{eq:rtag} becomes
\begin{align}
    v_{\rm in}(t)
    & =   A_{\rm r}  x(t - \tau_{\rm r}) e^{\jmath \theta_{\rm r}} +  \sum_{m=1}^M A_{{\rm h}, m} e^{\jmath\bar{\theta}_m} \notag \\
   &= A_{\rm r}  x(t - \tau_{\rm r}) e^{\jmath \theta_{\rm r}} +  A_{\rm h} e^{\jmath \theta_{\rm h}},
    \label{eq:signal_tag}
\end{align}
where $A_{{\rm h}, m} = \sqrt{2k_{\rm in} R_{\rm F}P_{\rm r}} h_{\rm u}(d_m)$. We distinguish here the RN contribution, $A_{\rm r}x(t-\tau_{\rm r}) e^{\jmath \theta_{\rm r}}$, and the HN contribution, $A_{\rm h} e^{\jmath \theta_{\rm h}} = \sum_{m=1}^M A_{{\rm h}, m} e^{\jmath\bar{\theta}_m}$.
Substituting \eqref{eq:signal_tag} into \eqref{eq:tag_output} gives the tag output as
\begin{align}
    v_{\rm out}(t) = \frac{R_{\rm H}}{R_{\rm F}} \beta v_{\rm in}^2(t)=\frac{R_{\rm H}}{R_{\rm F}} \beta \left(A_{\rm r}  x(t - \tau_{\rm r}) e^{\jmath \theta_{\rm r}} +  A_{\rm h} e^{\jmath \theta_{\rm h}} \right)^2, 
    \label{eq:vout_final} 
\end{align}
and inserting \eqref{eq:vout_final} into \eqref{eq:received_sig} finally yields
\begin{align}
    r(t)
    & =\eta \Big( A_{\rm r}^2  x^2(t - 2\tau_{\rm r}) e^{\jmath2\theta_{\rm r}} + 2A_{\rm r}A_{\rm h}  x(t - 2\tau_{\rm r})e^{\jmath(\theta_{\rm r}+\theta_{\rm h})}  \notag \\ 
     & \qquad \qquad \qquad \qquad \qquad+ A_{\rm h}^2 e^{\jmath2\theta_{\rm h}}\Big) e^{\jmath 2\theta_{\rm r}} + n(t).
     \label{eq:r_final}
\end{align}
The composition of the downlink signal \eqref{eq:r_final} captures the nonlinear action of the tag in its quadratic regime near the maximum range. The received signal now contains three terms:
\begin{enumerate}
    \item the first term (ranging signal only) corresponds to the output of a conventional HR system without helpers;
    \item the second term (intermodulation) carries the ranging waveform $x(t)$ with amplitude proportional to $2A_{\rm r}{A_{\rm h}}$, which is at least $6$~dB greater than in conventional HR, if $A_{\rm h} \geq A_{\rm r}$;
    \item the third term (helpers only) is unmodulated carrier.
\end{enumerate}
In helper-based HR, the receiver detects the intermodulation term (term 2 in the above) in order to gain the increase in signal to noise ratio (SNR) or range. Its sliding correlator or filter matched to $x(t)$ is unresponsive to the other two terms, since the first term contains
$x^2(t) = 1$ and the third term is a constant, neither of which correlate strongly against typical $x(t)$ sequences with good autocorrelation properties. With term 2 as the objective in \eqref{eq:r_final}, the main goal of the helper-based HR system design is maximizing the amplitude of the helper node contribution $A_{\rm h}$.

%% file: sec_TS.tex
Above, we saw that when the amplitude $A_{\rm h}$ of the HN contribution is greater than half of the ranging signal amplitude $A_{\rm r}$, helper-based system provides an increased SNR (or range), relative to conventional HR.  
The advantage of employing helper nodes is that the  increase of the tag output power is achieved by using simple constant-envelope signals, which are inexpensive and power-efficient.  Further, helper nodes can be added incrementally in order to provide the power boost needed for the desired SNR or range increase.
 Below, we evaluate how much improvement can be achieved with helper nodes in different operating schemes.

\subsection{Coherent Transmission Scheme}

A coherent transmission scheme ensures that HN signals arrive at the tag in phase, in spite of drifts in distances and in oscillator phases.  In Section~\ref{sec:PA}, we introduce an adaptive phase coherence method to achieve such a system. Here, we show that the benefits of coherent operation are substantial. Consider the amplitude $A_{\rm h}$ of the helper node contribution introduced in \eqref{eq:signal_tag}:
\begin{equation}
    A_{\rm h} \overset{\Delta}{=} \left|\sum_{m=1}^M A_{{\rm h}, m} e^{\jmath\bar{\theta}_m}\right|.
    \label{eq:magnitude_hn}
\end{equation}
If all $\bar{\theta}_m = \bar{\theta}$, then
\begin{equation}
    A_{\rm h} = \left|e^{\jmath\bar{\theta}} \sum_{m=1}^M A_{{\rm h}, m} \right| = \sum_{m=1}^M A_{{\rm h}, m} = M A_{\rm r},
\end{equation}
where the last equality assumes, for simplicity, that all helper tones arrive at the tag with the same amplitude as the ranging signal. Then, the power of the intermodulation term in \eqref{eq:r_final} is proportional\footnote{For notational convenience, here and in the following the signal power is assumed to be normalised to the link gain $\eta(d_{\rm r})$ defined below \eqref{eq:recieved_sig2}.} to
\begin{equation}
    P_{\rm i} = 4M^2A^4_{\rm r},
    \label{eq:power_interm_coh}
\end{equation}
which is $4M^2$ times the power of the ranging signal in a conventional HR system (first term in \eqref{eq:r_final}). As a result, the helper-based system provides an SNR boost of $20 \log (M) + 6$~dB, or about 18~dB for $M=4$ helpers. Since received signal power in HR varies as inverse sixth power of distance, a coherent helper system provides a range extension factor (REF) of
\begin{equation}
    \zeta_{\rm coh} = (4M^2)^{1/6} = \sqrt[3]{2M}.
    \label{eq:z_coh}
\end{equation}
This implies that $M=4$ helper nodes can double the range, compared to conventional HR. Even for the simplest case,i.e., a single helper ($M=1$) where coherence is no longer a consideration, \eqref{eq:z_coh} shows that REF is 1.26. A $26\%$ increase in range for almost no effort is appealing.

\subsection{Basic scheme -- Incoherent Transmission}
\label{sec:basicscheme}

Although the focus of this paper is ensuring that helper tones arrive coherently at the tag, there may be a role for a simpler incoherent system, one which omits the adaptive phase coherence algorithm, as in \cite{lav2019on}. 
According to \eqref{eq:magnitude_hn}, if the phases $\bar{\theta}_m$ are mutually independent and uniformly distributed in $[-\pi, \pi)$, then the mean helper node power at the tag is the power-wise sum 
\begin{equation}
    \bar{P}_{\rm h} = \sum_{m=1}^MA_{{\rm h}, m}^2 =MA_{\rm r}^2,
\end{equation}
where the last equality, for simplicity, sets all helper amplitudes equal to $A_{\rm r}$. 
The mean power of the intermodulation term then becomes
\begin{equation}
    \bar{P}_{\rm i} = 4\bar{P}_{\rm h} A_{\rm r}^2 = 4 M A_{\rm r}^4.
    \label{eq:power_interm_mean}
\end{equation}
This is $4M$ times the power of the signal term in a conventional HR system. On the other hand, this average is $M$ times smaller than the constant power \eqref{eq:power_interm_coh} of the coherent transmission. At the mean power of the intermodulation term \eqref{eq:power_interm_mean}, the SNR boost is then $10\log(M)+6$ dB, equating to 12dB for $M=4$ with a corresponding REF of $1.58$.

What about the instantaneous power $P_{\rm i} = 4A_{\rm h}^2 A_{\rm r}^2$? From \eqref{eq:magnitude_hn}, the magnitude $A_{\rm h}$ varies as the individual phases drift with changing distances to the tag and different individual LO phase offsets. 
Suppose that there are many helpers. Then, invoking the central limit theorem  we can model the real and imaginary components of the helper node contribution  as Gaussian variables with weak mutual dependence. It is well known  that the resulting complex Gaussian process has a Rayleigh distributed amplitude ($A_{\rm h}$) and an
exponentially distributed power ($P_{\rm h} = A_{\rm h}^2$) \cite{cavers2006mobile}. 
Consequently, the cumulative
distribution function (CDF) of $P_{\rm i}$ for many helpers is asymptotically
\begin{equation}
    \text{Pr}[P_{\rm i} \leq  z] = 1-e^{- \displaystyle \frac{z}{\bar{P}_{\rm i} }} = 1-e^{- \displaystyle \frac{z}{4 M A_{\rm r}^4} }.
    \label{eq:prob_nonch}
\end{equation}
From \eqref{eq:prob_nonch}, the instantaneous power of the helper-based HR system with incoherent transmissions is  $4M$ times the power of conventional HR approximately $37\%$ of the
time, and  when the phases are almost aligned it can occasionally  approach the power of coherent transmission, which is $4M^2$ times that of conventional HR. However, infrequent
deep fades can take it \textit{below} the power of conventional HR, with probability $1-\exp(-1/4M) \approx 1/4M$ ($6.1\%$ for four helpers).
If the range is extended by a factor $\zeta_{\rm inc}$, then the dropout probability becomes $1-\exp(-\zeta_{\rm inc}^6/4M)$ which grows rapidly. For such a system, range extension may not
be an appropriate measure. Instead, if this incoherent system operates at the same range as conventional HR, it provides much higher SNR on average, at the cost of dropouts a fraction $1/(4M)$ of the time. Therefore, applications that can tolerate infrequent dropouts may well find incoherent transmission attractive.

\subsection{Brute Force Conventional HR}
\label{sec:brute_force}

A helper-based system invests the power of the ranging node and $M$ helpers into SNR
improvement or range increase. Assume, again for simplicity, that all helpers have the same
transmit power as the ranging node and that all path losses are equal. A fair comparison might then allow conventional HR to increase its transmit power by a factor of $M+1$. With this brute-force improvement, the received power at the tag in conventional HR is
\begin{equation}
    P_{\rm r} = (M+1)^2A_{\rm r}^4.
    \label{eq:convent_power}
\end{equation}
%;
Comparison with \eqref{eq:power_interm_coh} and consideration of the inverse sixth power dependence on range gives the REF as
\begin{equation}
    \zeta_{\rm conv} = \sqrt[\displaystyle 3]{M+1}.
    \label{eq:ref_conv}
\end{equation}
For a single helper ($M=1$), therefore, SNR and REF of the conventional HR system are as good as those of the coherent helper-based system \eqref{eq:power_interm_coh}. However, conventional HR rapidly falls behind the coherent helper system as the number of helper nodes increases, leaving the coherent system with an asymptotic $26\%$ greater range.
To match the range of a helper-based system, the conventional system must increase the transmit power of its RN, which also increases the weight and cost of the amplifier and power supply.
It is worth noting here that HR systems have increased linearity requirements on the RN in order to minimize the parasitic 2nd harmonic leakage, leading to the need for more carefully designed amplifiers and, depending on the modulation, implementation of additional linearization techniques \cite{gallagher2014linearization, lav2019parasitic}. In contrast, the helper-node approach is distributed and incremental, so that helpers can be added only as needed.  Also, the set of helper nodes boosts the effective power of every RN, in case there is more than one.

%% file: sec_PA.tex
\subsection{Two-mode Transmission with Phase Adjustment}
\label{sec:transm_schm}

\begin{figure}[t!]
  \vspace{-\baselineskip}
              \centering
   \includegraphics[width =0.85\linewidth]{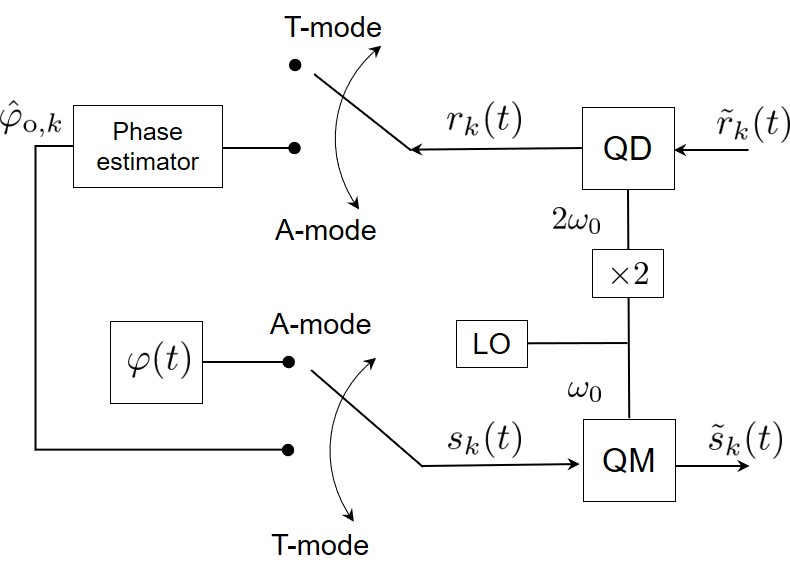}
   \caption{\small{
   Block scheme of a helper node that can operate in one of two modes: a phase adjustment mode (A) or a fixed-phase transmission mode (T). QM and QD here stand for  the  quadrature modulator and demodulator, respectively.}}
%   \label{fig:adaptive}
       \vspace*{-0.3cm}
        \label{fig:adaptive}
\end{figure}

From Section~\ref{sec:DS}, helper nodes transmit a fixed-phase tone, a mode which we will call T-mode
(tone). However, from \eqref{eq:power_interm_coh} and \eqref{eq:z_coh}, we want the individual helper node tones to combine
coherently at the tag, in order to maximize the tag's output power.
That requires an additional
mode (A, for adjustment) in which a helper node adjusts its transmit phase $\theta_m$ in response to the tag output signal $\tilde{v}_{\rm out}(t)$, which implies that helper nodes must also be able to receive signals on the downlink. Fig.~\ref{fig:adaptive} outlines the helper node structure that supports this two-mode operation.
When in mode A, the helper node sweeps its phase around the circle as $\theta_m + \varphi(t)$, where $\varphi(t) = 2 \pi {t}/{T_{\rm s}}, 0\leq  t \leq T_{\rm s}$ and $T_{\rm s}$ is the phase sweep duration. During this time, it processes the received response from the tag in order to calculate the
new value of transmit phase to use when it next returns to T mode. 

\begin{figure}[t!]
    \centering
    \includegraphics[width=0.8\columnwidth]{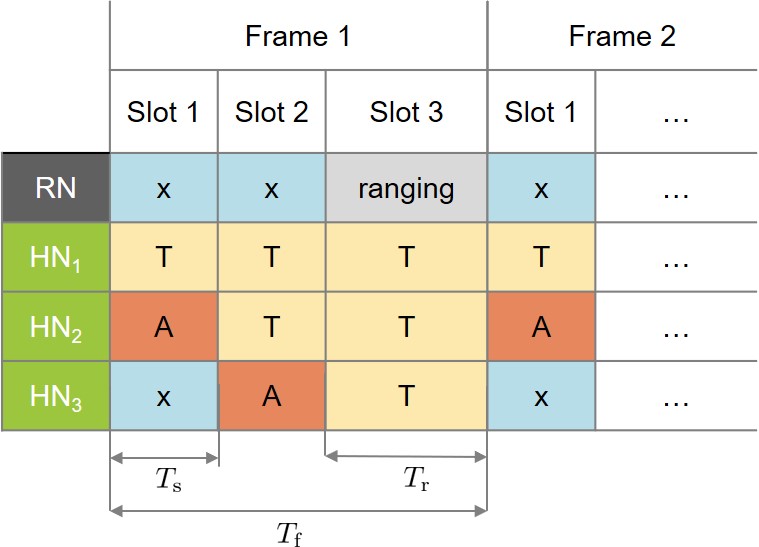}
    \caption{\small{Simplified transmission frame structure for $M=3$ helper nodes. The frame $T_{\rm f}$ is organized in $M$ time slots: $M-1$ phase adjustment slots of duration $T_{\rm s}$ each followed by a $T_{\rm r}$-long slot for ranging, i.e., $T_{\rm f} = (M-1)T_{\rm s} + T_{\rm r}$.
    }}
       \label{fig:Schdl}
       \vspace*{-0.3cm}
\end{figure}

To coordinate the change between the two different modes at each helper and the operation of the ranging node, the transmission time consists of a series of frames. Each frame begins with a phase adjustment interval of $M-1$ time slots, each of duration $T_{\rm s}$, used for helper phase adjustments, followed by a ranging interval of duration $T_{\rm r}$, resulting in a total frame duration of $T_{\rm f} = (M-1)T_{\rm s} +T_{\rm r}$. Furthermore, during phase adjustment slots only a single helper is in mode A at any given time. Fig.~\ref{fig:Schdl} exemplifies a simple frame structure for $M=3$ helpers. In  slot 1, helper 1, which is in mode T, sends a tone and helper 2, in mode A, sends a phase sweep that allows it to align its phase with that of helper 1. In slot 2, helpers 1 and 2 send their now-aligned tones in mode T, while helper 3, in mode A,  sends the phase sweep in order to align its phase with that of the sum of helpers 1 and 2. The adjustment interval continues to add one helper at a time, slot by slot, until all $M$ helpers are aligned. In the absence of receiver noise, $M-1$ phase adjustment slots are sufficient here to achieve exact phase coherence.  During the ranging interval that follows (slot 3 in Fig.~\ref{fig:Schdl}), the RN sends the ranging signal while the helpers, all now in mode T, send tones which arrive coherently at the tag, thereby boosting the amplitude of the intermodulation term at the tag output.

In more detail, the signal arriving at the tag in phase adjustment slot $i$, $i = 1, \ldots, M-1$, is composed of the phase-sweep of helper $i+1$ and the fixed-phase tones of helpers $1$ to $i$, so the complex envelope of the tag input is
\begin{align}
\label{eq:vin_a}
    v_{\rm in}(t-(i-1)T_{\rm s}) &= A_{{\rm h}, i+1} e^{\jmath \bar{{\theta}}_{i+1}}e^{\jmath \varphi(t-(i-1)T_{\rm s})} + \sum_{m=1}^i  A_{{\rm h}, m} e^{\jmath \bar{\theta}_{m}} \notag \\
    & = A_{{\rm h}, i+1} e^{\jmath \bar{\theta}_{i+1}}e^{\jmath \varphi(t-(i-1)T_{\rm s})} + A_{{\rm p}, i} e^{\jmath \theta_{{\rm p},i}},
\end{align}
where $A_{{\rm p}, i} e^{\jmath \theta_{{\rm p},i}} = \sum_{m=1}^i  A_{{\rm h}, m} e^{\jmath \bar{\theta}_{m}}$ is the slot-$i$ partial sum of helper node signals, since only a subset of helpers is in T mode during any phase adjustment slot. After the final adjustment in slot $M-1$, the sum is no longer partial so that $A_{\rm h}e^{\jmath \theta_{\rm h}} = A_{{\rm p}, M} e^{\jmath \theta_{{\rm p},M}}$.  Moving to the downlink, \eqref{eq:vout_final} gives the second-harmonic complex envelope
at the tag output as
\begin{align}
    \label{eq:vout_a}
    v_{\rm out}(t-(i-1)T_{\rm s}) & = \frac{R_{\rm H}}{R_{\rm F}} \beta {v}_{\rm in}^2(t-(i-1)T_{\rm s})  \\
   &=  \frac{R_{\rm H}}{R_{\rm F}} \beta
  \left( {A}_{{\rm h}, i+1} e^{\jmath\left(\bar{\theta}_{i+1}+\varphi(t-(i-1)T_{\rm s})\right)}+ A_{{\rm p},i} e^{\jmath \theta_{{\rm p},i}} \right)^2 \notag
\end{align}
and \eqref{eq:r_final} then gives the second-harmonic receiver input at helper $i+1$, which is in mode A, as
\begin{align}
    \label{eq:baseband_recieved_kth}
    r_{i+1}(t-(i-1)T_{\rm s})
    & = 
    \eta_{i+1} \Big( A_{{\rm h}, i+1}^2e^{\jmath 2\bar{\theta}_{i+1}}e^{\jmath 2\varphi(t-(i-1)T_{\rm s})}  \\
    &  \quad \qquad +2A_{{\rm h}, i+1}A_{{\rm p}, i}e^{\jmath (\bar{\theta}_{i+1}+ \theta_{{\rm p},i})}e^{\jmath \varphi(t-(i-1)T_{\rm s})} \notag \\
     &  \quad \qquad + A_{{\rm p}, i}^2e^{\jmath 2{\theta}_{{\rm p},i}}\Big)e^{-\jmath 2\omega_0\tau_{i+1}} +n_{i+1}(t), \notag
\end{align}
 where $\eta_{i} = h_{\rm d}(d_i) \beta \sqrt{k_{\rm out}R_{\rm H}R_{\rm rx}}/{R_{\rm F}}$. 
For simplicity, in \eqref{eq:vin_a}--\eqref{eq:baseband_recieved_kth} we ignore the propagation delay in $\varphi(t)$, on the grounds that $\tau_{i+1} \ll T_{\rm s}$, but we revisit this point later.
From \eqref{eq:vout_a}, the partial sum $A_{{\rm p},i}e^{\jmath \theta_{{\rm p}, i}}$ in adjustment slot $i$ reflects prior adjustments, so that helper $i+1$ should change its phase to equal that of the partial sum, $\theta_{{\rm p}, i}$.   In the absence
of error, this maximizes the amplitude of the next partial sum so that $A_{{\rm p}, i+1} = A_{{\rm h}, i+1} +A_{{\rm p}, i}$.
According to \eqref{eq:baseband_recieved_kth}, the phase offset required to achieve this is 
\begin{equation}
     \varphi_{{\rm o}, i+1} = \theta_{{\rm p}, i}-\bar{\theta}_{i+1} = \theta_{{\rm p}, i} - \theta_{i+1} +\omega_0\tau_{i+1}.
     \label{eq:phase_offcet}
\end{equation}
Below, we describe a method to estimate $\varphi_{{\rm o}, i+1}$ from the received signal $r_{i+1}(t-(i-1)T_{\rm s})$.

\subsection{Phase Offset Estimation}
\label{sec:phase_estm}

Consider \eqref{eq:baseband_recieved_kth} and note that the individual terms that constitute it are separable with respect to $\varphi(t)$ via the following integrals
\begin{align}
    {G}_{0, i} &= \overset{{iT_{\rm s}}}{\underset{{(i-1)T_{\rm s}}}{\int}} r_{i+1}(t-(i-1)T_{\rm s}) \text{d}t \label{eq:Fintegral1} \\
    &=  \underbrace{\eta_{i+1} T_{\rm s} A^2_{{\rm p}, i}e^{\jmath2\left(\theta_{{\rm p}, i}-\omega_0\tau_{i+1}\right)}}_{\dot{G}_{0, i}} + \int_0^{T_{\rm s}} n(t) \text{d}t  = \dot{G}_{0, i} + n_{0, i}, \notag
\end{align}
    \begin{align}
    {G}_{1, i} & = \overset{{iT_{\rm s}}}{\underset{{(i-1)T_{\rm s}}}{\int}} r_{i+1}(t-(i-1)T_{\rm s}) e^{-\jmath 2 \pi \frac{t-(i-1)T_{\rm s}}{T_{\rm s}}} \text{d}t \label{eq:Fintegral2} \\
    &= \underbrace{2 \eta_{i+1} T_{\rm s} A_{{\rm h}, i+1}A_{{\rm p},i} e^{\jmath\left(\bar{\theta}_{i+1}+\theta_{{\rm p}, i} -2\omega_0\tau_{i+1} \right)}}_{\dot{G}_{1, i}} \notag \\
     & \qquad \qquad \qquad \qquad \qquad +  \int_0^{T_{\rm s}}n(t) e^{-\jmath 2 \pi \frac{t}{T_{\rm s}}} \text{d}t \notag \\
    &= \dot{G}_{1, i} +n_{1, i}. \notag  \\
   {G}_{2, i} & = \overset{{iT_{\rm s}}}{\underset{{(i-1)T_{\rm s}}}{\int}} r_{i+1}(t-(i-1)T_{\rm s}) e^{-\jmath 4 \pi \frac{t-(i-1)T_{\rm s}}{T_{\rm s}}} \text{d}t \label{eq:Fintegral3} \\
    & = \underbrace{\eta_{i+1} T_{\rm s} A^2_{{\rm h}, i+1}e^{\jmath2\left(\bar{\theta}_{i+1}-\omega_0\tau_{i+1}\right )}}_{\dot{G}_{2, i}} + \int_{0}^{T_{\rm s}} n(t) e^{-\jmath 4 \pi \frac{t}{T_{\rm s}}} \text{d}t \notag \\
    &= \dot{G}_{2, i} +n_{2, i},\notag 
\end{align}
where $n_{k, i} = \int_0^{T_{\rm s}}n(t) e^{-\jmath 2 \pi \frac{(k-1)}{T_{\rm s}}t} \text{d}t$ and the overhead dot denotes the noise-free component. 
Since the basis functions $\exp(-\jmath 2 \pi \frac{k}{T_{\rm s}}t)$ in \eqref{eq:Fintegral1}-\eqref{eq:Fintegral3} are orthogonal and have equal energy, the noise terms $n_{0, i}$, $n_{1, i}$, $n_{2, i}$ are independent and have equal variance $\sigma_n^2 = T_{\rm s} N_0$.  For each of the integration outputs, we can define individual SNRs
\begin{align}
\label{eq:gamma0}
 \gamma_{0, i} &= \frac{|\dot{G}_{0, i}|^2}{\sigma_n^2} =  \frac{\eta_{i+1}^2 A^4_{{\rm p}, i}T_{\rm s}}{N_0}, \\
        \gamma_{1, i} & = \frac{|\dot{G}_{1, i}|^2}{\sigma_n^2} =  \frac{ 4 \eta_{i+1}^2 A_{{\rm h}, i+1}^2A^2_{{\rm p},i}T_{\rm s}}{N_0}, \label{eq:gamma1} %= 4 \left(\frac{A_{{\rm h}, k}}{A_{\rm p}}\right)^2 \gamma_0.
        \\
            \gamma_{2, i}  &= \frac{|\dot{G}_{2, i}|^2}{\sigma_n^2} =  \frac{\eta_{i+1}^2 A_{{\rm h}, i+1}^4 T_{\rm s}}{N_0}.
     \label{eq:gamma2}
\end{align}
In the above, both $\eta_{i+1}^2$ and $N_0$ are proportional to $R_{\rm rx}$. Consequently, the SNRs do not depend on the Rx antenna resistance.
Using \eqref{eq:Fintegral1}-\eqref{eq:Fintegral3} the phase offset $\varphi_{{\rm o}, i+1}$ can then be computed as
\begin{equation}
    \hat{\varphi}_{{\rm o}, i+1} =  \arg({G}_{0,i}{G}_{1,i}^*), 
    \label{eq:phase_estim}
\end{equation}
where $(\cdot)^{*}$ denotes complex conjugate.  
Note that while we do not directly use the output $G_{2, i}$ of the third integrator for phase estimation, it provides a useful SNR reference that is independent of the partial set amplitude $A^4_{{\rm p}, i}$, so that, expressing \eqref{eq:gamma0}--\eqref{eq:gamma1} via \eqref{eq:gamma2}, we obtain
\begin{align}
\label{eq:gamma_0i}
 \gamma_{0, i} &= \frac{A^4_{{\rm p}, i}}{A^4_{{\rm h}, i+1}} \gamma_{2, i}, \\
        \gamma_{1, i} & = 4 \frac{A^2_{{\rm p}, i}}{A^2_{{\rm h}, i+1}} \gamma_{2, i}.
        \label{eq:gamma_1i}
\end{align}

Phase estimation in \eqref{eq:phase_estim} is similar to a phase error detection in a phase-locked loop. In contrast to operating directly on the magnitude of received measurements, it first projects the received signal onto a set of orthogonal subspaces and then extracts the phase. This separates the individual signal components and, by averaging over the sweep time, reduces the effect of additive noise.

%% file: sec_AN.tex
Above, we presented a transmission scheme and a phase estimation method to phase-align helper node signals for coherent signal combining at the tag. In the absence of noise, it achieves phase alignment in $M-1$ slots. 
When an additive noise is present, how well the phases can be aligned depends on the phase estimation accuracy of \eqref{eq:phase_estim}. The following section presents an analysis of the phase adjustment accuracy of the proposed approach and its effect on the REF of the helper-based system. 

\subsection{Phase Estimation}
Consider \eqref{eq:phase_estim} and rewrite ${G}_{0, i}$, ${G}_{1, i}$ in terms of $\gamma_{0, i}, \gamma_{1, i}$
\begin{align}
    {G}_{0, i}{G}_{1, i}^* = \sigma_n^2 \sqrt{\gamma_{0, i} \gamma_{1, i}}e^{\jmath \varphi_{{\rm o}, i+1}} + n_{{\rm t}, i},
    \label{eq:g0g1}
\end{align}
where
\begin{equation}
     n_{{\rm t}, i} = \sigma_n\sqrt{\gamma_{1, i}} e^{-\jmath\arg(\dot{G}_{1, i})} n_{0, i}+\sigma_n\sqrt{\gamma_{0, i}} e^{-\jmath\arg(\dot{G}_{0, i})}n_{1, i} + n_{0, i}n_{1, i}^*.
     \label{eq:noise}
\end{equation}
From \eqref{eq:g0g1}, the phase offset $\varphi_{{\rm o}, i+1}$ can be written as
\begin{equation}
    \hat{\varphi}_{{\rm o},i+1} = \arg({G}_{0, i}{G}_{1, i}^*) = \varphi_{{\rm o}, i+1} + \varphi_{{\rm er}, i},
    \label{eq:phase_off_est}
\end{equation}
where $\varphi_{{\rm er}, i}$ is the noise-induced phase estimation error in the $i$-th phase adjustment slot.

The three terms that constitute $n_{{\rm t}, i}$ in \eqref{eq:noise} are uncorrelated, so its variance is
\begin{equation}
    \sigma_{i}^2 = \left( \gamma_{0, i}+ \gamma_{1, i}+1 \right) \sigma_n^4.
\end{equation}
Because of the quadratic term $n_{0, i} n_{1, i}^*$  the noise $n_{{\rm t}, i}$ is not strictly Gaussian. Nevertheless, when its two linear components dominate the quadratic term ($\gamma_{0, i} + \gamma_{1,i} \geq 1$), $n_{{\rm t}, i}$  can be well approximated as zero-mean complex Gaussian. In this case, ${G}_{0,i}{G}_{1, i}^*$  has a non-central complex Gaussian distribution with a K-factor (ratio of the squared mean to the variance, another SNR measure) equal to
\begin{equation}
\label{eq:K_ifactor}
    K_i  \overset{\Delta}{=} \frac{\left(\sigma_n^2 \sqrt{\gamma_{0, i} \gamma_{1, i}}\right)^2}{\left( \gamma_{0, i}+ \gamma_{1, i}+1 \right) \sigma_n^4} = \frac{\gamma_{0,i} \gamma_{1,i}}{ \gamma_{0, i}+ \gamma_{1, i}+1}. %\approx \frac{\gamma_0 \gamma_1}{\gamma_0+ \gamma_1},
\end{equation}
As for the phase error $\varphi_{{\rm er}, i}$, its distribution is that of the phase of \eqref{eq:g0g1} when $\varphi_{{\rm o},i+1} = 0$. From \cite[eq.~(5.33)]{molisch2012wireless}, its PDF is
\begin{align}
\label{eq:phase_pdf}
    f_{\varphi_{{\rm er}, i}}(\varphi_{{\rm er}, i}, K_i) =  \frac{e^{-K_i}}{2\pi} \Big( 1+ \sqrt{4\pi K_i} x e^{{K_i}x^2} Q(-\sqrt{2K_i}x )\Big),
\end{align}
where $x = \cos \varphi_{{\rm er}, i}$ and $Q(\cdot)$ denotes the complementary cumulative distribution function (CCDF) of a standard Gaussian distribution.

\subsection{Phase Adjustment}
\label{sec:phase_adjtsm_acc}

Consider now the phase adjustment process exemplified by Frame 1 in Fig.~\ref{fig:Schdl}. 
Suppose, for simplicity, that all signals arrive at the tag with amplitudes equal to that of the ranging node $A_{\rm r}$, so that $A_{{\rm h}, m} = A_{\rm r}$ and $\eta_m = \eta, m = 1, \ldots M$. Our primary interest then is how the partial helper set amplitude $A_{{\rm p}, i}$ changes from slot to slot with respect to $A_{\rm r}$. To evaluate this, we introduce the amplitude ratio $\alpha_i$, defined as
\begin{equation}
    \alpha_i = \frac{A_{{\rm p}, i}}{A_{\rm r}}.
    \label{eq:alpha_n}
\end{equation}
In the first phase adjustment slot when $i=1$, we have that $A_{{\rm p}, 1} = A_{{\rm h}, 1} =  A_{\rm r}$ so that $\alpha_1 = 1$. 
In the second slot ($i=2$), we obtain from elementary geometry that
\begin{equation}
    A_{{\rm p}, 2} = \sqrt{ A^2_{\rm r}+2A_{{\rm p}, 1}A_{\rm r} \cos \varphi_{{\rm er}, 1} + A_{{\rm p}, 1}^2} = \alpha_2 A_{\rm r} 
    \label{eq:first_slot}
\end{equation}
where $\alpha_2 = \sqrt{ 1+2\alpha_1 \cos \varphi_{{\rm er}, 1} + \alpha_1^2} \in [0, 2]$. 
Generally, $A_{{\rm p}, i} = \alpha_{i} A_{\rm r}$ where 
\begin{equation}
    \alpha_i = \sqrt{1+2\alpha_{i-1}\cos(\varphi_{{\rm er},i-1})+\alpha_{i-1}^2},
    \label{eq:alpha}
\end{equation}
 so that $\alpha_i \in [0, i]$ for any $2 \leq i \leq M$. Combining \eqref{eq:K_ifactor} with \eqref{eq:gamma_0i}-\eqref{eq:gamma_1i}, the K-factor in the adjustment slot $i$ becomes
\begin{equation}
     K_i = \frac{4 \alpha_{i}^6 \gamma_2^2}{\alpha_{i}^4\gamma_2+4\alpha_{i}^2\gamma_2+1}, %\approx \frac{4\alpha_{n-1}^4}{\alpha_{n-1}^2+4} \gamma_2. 
     \label{eq:Kn}
\end{equation}
where $\gamma_2 =\eta^2  T_{\rm s} A_{\rm r}^4/N_0$.

From \eqref{eq:alpha}, the amplitude ratio $\alpha_i$ for $i>1$ is a random variable that depends both on the phase estimation error  and the amplitude ratio in the previous slot.
%$\alpha_i$
In order to derive its distribution, we first compute a conditional PDF $f(\alpha_i| \alpha_{i-1})$. Assuming that $\varphi_{{\rm er}, i}$ ($1 \leq i \leq M-1$) are independent random variables distributed according to \eqref{eq:phase_pdf}, we can use \eqref{eq:alpha} and a change of variables \cite{davenport1958introduction} to compute 
\begin{align}
   \label{eq:PDF_x}
    f_{\alpha_i}(\alpha_i| \alpha_{i-1}) = &\frac{e^{-K_{i-1}}}{\pi} \frac{\alpha_i}{\alpha_{i-1}} \frac{1}{\sqrt{1-z^2}} \\
    &\qquad \times \Big( 1+ \sqrt{4\pi K_{i-1}} z e^{K_{i-1}z^2} Q(-\sqrt{2K_{i-1}}z )\Big) ,
 \notag
\end{align}
where $z = \left(\alpha_i^2-\alpha_{i-1}^2-1\right)/{2\alpha_{i-1}}$.
Note that although it is not shown explicitly in \eqref{eq:PDF_x}, $K_{i-1}$ depends on $\alpha_{i-1}$ via \eqref{eq:Kn}. Expression \eqref{eq:PDF_x} highlights the Markov nature of the change of the amplitude ratio from slot to slot.
Marginalizing it over the distribution of $\alpha_{i-1}$, we obtain the following recursive formula for the PDF of $\alpha_i$
 \begin{align}
    f_{\alpha_i}(\alpha_i) &= \int_0^{i-1}  f(\alpha_i| \alpha_{i-1})  f(\alpha_{i-1})\text{d}\alpha_{i-1}.
    \label{eq:alpha_pdf}
\end{align}
 The closed-form solution to \eqref{eq:alpha_pdf} can be obtained for $\alpha_2$ as
 \begin{equation}
     f_{\alpha_2}(\alpha_2) = \frac{e^{-K_{1}}}{\pi}  \sqrt{\frac{4}{4-\alpha_2^2}} \Big( 1+ \sqrt{\pi K_{1}} z e^{K_{1}z^2/4} Q(-\sqrt{0.5K_{1}}z )\Big),
     \label{eq:final_a2}
 \end{equation}
 where $z = {\alpha_2^2-2}$.
 For any $i>2$, $f_{\alpha_i}(\alpha_i)$ can subsequently be found from \eqref{eq:alpha_pdf} by applying numerical methods\footnote{Note that with minor modifications the presented analysis also holds for a more general case of unequal helper node amplitudes.}. %which is described in App. B. 

  Fig.~\ref{fig:PDFs} illustrates \eqref{eq:alpha_pdf} by showing $f_{\alpha_i}(\alpha_i)$ for different values of $i$ and input SNR $\gamma_2$. 
 In Fig.~\ref{fig:Kappa_PDF_2}, we observe that the theoretical PDFs computed via \eqref{eq:final_a2} coincide well with the empirical ones for both considered SNRs. Fig.~\ref{fig:Kappa_PDF} on the other hand, illustrates that, as the phase adjustment proceeds from slot to slot  and the number of helpers $i$ grows, so does the likelihood that the partial set amplitude $A_{{\rm p}, i}$ increases, as higher values of the amplitude ratio in previous slots lead to improved phase estimation performance  (via increased K-factor) and thus better chances for obtaining higher amplitude ratio and higher SNR in the current slot. Further, the long-tailed asymmetry of the PDFs shows that, while $A_{{\rm p},i}$ is almost always close to $i A_{\rm r}$, it can fall to very low values on rare occasions. Therefore, our design analysis should be based on performance percentiles, rather than on the average values.

   \begin{figure}[t!]
    \centering
    \begin{subfigure}[b]{0.9\linewidth}
    \input{figures/Kappa_PDF2.tex}
   % \vspace*{-0.4cm}
    \caption{\small{}}   
   \label{fig:Kappa_PDF_2}
\end{subfigure}
   \begin{subfigure}[b]{0.9\linewidth}
    \centering
    \input{figures/PDF_alphan.tex}
    \caption{\small{}}   
   \label{fig:Kappa_PDF}
   \end{subfigure}
   \caption{\small{Exemplary PDFs of $\alpha_i$ for (a) $i=2$ and (b) $i \in [2, 5]$.}}
      \vspace*{-0.3cm}
      \label{fig:PDFs}
\end{figure}
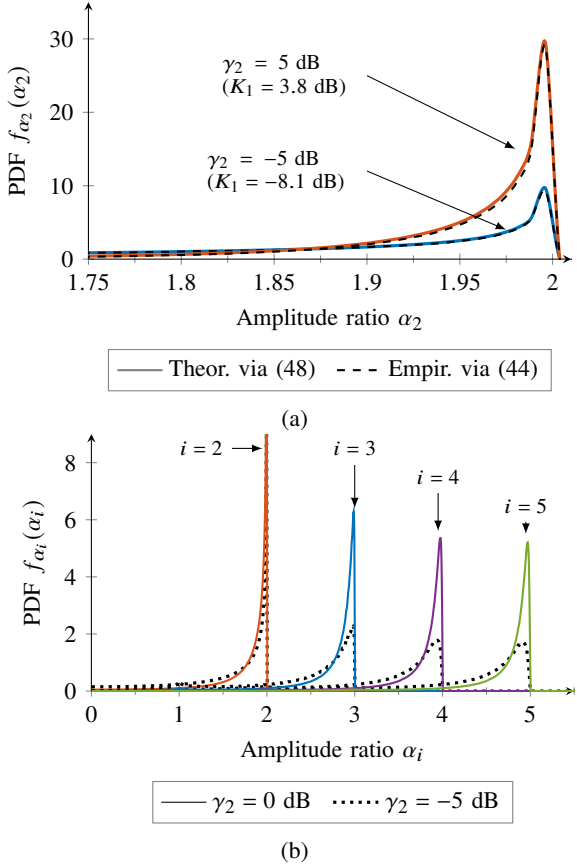 

\begin{figure}[t!]
    \centering
   \input{figures/REF_prcnt.tex}
         \vspace*{-0.1cm}
\caption{\small{REF percentiles as a function of the number of helpers $M$ for input SNR $\gamma_2 = - 5$~dB. The solid black line indicates  REF of the conventional brute-force HR system without helper nodes $\zeta_{\rm conv} = \sqrt[3]{M+1}$ \eqref{eq:ref_conv} while the dashed line shows REF of a fully coherent helper-based system $\zeta_{\rm coh} = \sqrt[3]{2M}$ \eqref{eq:z_coh}.}}
   \label{fig:ref_prcnt}
   \vspace*{-0.3cm}
\end{figure}
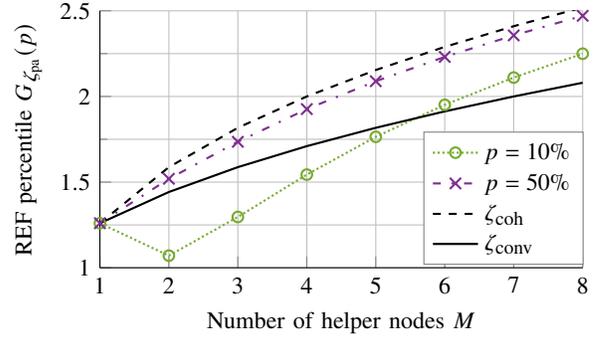

\subsection{Effect on Ranging SNR and REF}
Now we demonstrate the range extension provided by our helper-based system and compare it to the range extension achievable by a conventional system that is given the same total transmit power.  The helper-based HR system uses the intermodulation term (the second term of \eqref{eq:r_final}) for ranging. Previously, we saw that its instantaneous power is 
\begin{equation}
    P_{\rm i} = 4  A_{\rm h}^2 A_{\rm r}^2 \leq 4A_{\rm r}^2 \left(\sum_{m=1}^M A_{{\rm h}, m}\right)^2,
    \label{eq:power_inst}
\end{equation}
 with the upper limit being achieved when all $M$ helper tones perfectly phase-align at the tag. Suppose again that $A_{{\rm h}, m} = A_{\rm r}$ for all $m\leq M$ and consider \eqref{eq:power_inst}. We can express it as 
\begin{equation}
    P_{\rm i} =  4 \alpha_{M}^2 A_{\rm r}^4,
    \label{eq:ranging_pwr}
\end{equation}
where $\alpha_{M} = A_{\rm h}/A_{\rm r} \in [0, M]$ is the amplitude ratio after the last, $(M-1)$-th, phase adjustment slot. Comparing \eqref{eq:ranging_pwr} with the power of the ranging term (term 1 in \eqref{eq:r_final}), we see that the SNR boost is proportional to $4\alpha_M^2$ while the corresponding REF for phase adjustment is then $\zeta_{\rm pa} = \sqrt[3]{2 \alpha_{M}}$. With another change of variable, we obtain the PDF of $\zeta_{\rm pa}$ as
\begin{align}
    f_{\zeta_{\rm pa}} (\zeta_{\rm pa}) &= 1.5\zeta^2_{\rm pa}  f_{\alpha_M} (\zeta_{\rm pa}).
    \label{eq:pdf_ref}
\end{align}
Its CDF is $F_{\zeta_{\rm pa}}(\zeta_{\rm pa})=\int_0^{\zeta_{\rm pa}} f_{\zeta_{\rm pa}} (\zeta_{\rm pa}) \mathrm{d}\zeta_{\rm pa}$. We also define the inverse function of the CDF as
\begin{equation}
    G_{\zeta_{\rm pa}}(p) \overset{\Delta}{=} \zeta_{\rm pa}:  F_{\zeta_{\rm pa}} (\zeta_{\rm pa}) = 0.01p.
\end{equation}
That is, $G_{\zeta_{\rm pa}}(p)$ is the value of the REF $\zeta_{\rm pa}$ at the $p$-th percentile. Note that since $\zeta_{\rm pa} = \sqrt[3]{2 \alpha_{M}}$, $G_{\zeta_{\rm pa}}(p)$ depends both on the input SNR $\gamma_2$ and the number of helpers $M$.
Fig.~\ref{fig:ref_prcnt} depicts $G_{\zeta_{\rm pa}}(p)$ as a function of $M$ for input SNR $\gamma_2 = - 5$~dB. It shows a conservative 10th
percentile as a worst acceptable case, as well as the more optimistic 50th  percentile along with the corresponding REFs of a brute-force conventional HR system $\zeta_{\rm conv} = \sqrt[3]{M+1}$ \eqref{eq:ref_conv} and a fully coherent helper-based system $\zeta_{\rm coh} = \sqrt[3]{2M}$ \eqref{eq:z_coh}. All three systems have the same total transmitted power.
We observe from Fig.~\ref{fig:ref_prcnt} that, at this SNR ($\gamma_2  = -5$ dB), one could expect a range extension factor greater than one ($\zeta_{\rm pa}>1$) even in the worst-case scenario\footnote{At lower SNRs the long tails of $f_{\alpha_M}(\alpha_M)$ can cause the $10$th percentile REF for $M>1$ to fall below that of the constant REF provided by a single helper, which we observe in Fig.~\ref{fig:ref_prcnt} for $M=2$. Nevertheless, adding more  helpers consistently improves the $10$th percentile performance }, while in $50\%$ of cases the REF of the helper-based system well exceeds that of the conventional brute-force approach when both systems are given the same total transmission power.

\vspace{-0.2cm}
\subsection{Slot duration limitations}
\label{sec:SLD}

In our helper-based system, the phase adjustment interval consists of $M-1$ slots followed by a ranging interval. Each phase adjustment slot lasts $T_{\rm s}$ seconds during which the A-mode helper sweeps its phase according to the sweep function $\varphi(t) = 2 \pi t/T_{\rm s}$ and collects the tag response in order to compute its phase correction according to \eqref{eq:phase_estim}. While it appears that the slot duration $T_{\rm s}$ is a free parameter of choice, there are several considerations that limit its practical range.

The $i$-th phase adjustment slot ($i>1$) is launched with the helper node phases of nodes $1$ to $i$ obtained by the end of the preceding slot. The primary concern then is the phase drift due to frequency errors in LOs of the helpers and, if they or the tag are moving, differences in their Doppler shifts. Phase drift causes increasing de-coherence among helpers across the slot time, with a corresponding slow reduction of the amplitude ratio $\alpha_i$ through the shrinking of the $A_{{\rm p}, i}$ factor.
Suppose that due to LO and movement-induced frequency shifts the $k$-th helper signal arriving at the tag has a frequency
error $\omega_{{\rm er}, k}$ at the end of the phase adjustment slot $T_{\rm s}$, which causes a growing phase difference between the different helper tones. The longer the slot duration, the greater the phase difference. Since the phase difference must be contained over all $M-1$ phase adjustment slots, we require that
\begin{equation}
    T_{\rm s} \leq \frac{\Delta \theta_{\rm max}}{(M-1)\omega_{{\rm er}, \rm max}},
    \label{eq: freq_drift_limit}
\end{equation}
where $\Delta \theta_{\rm max}$ is the maximum allowable phase difference and $\omega_{{\rm er}, \rm max}$ is the maximum frequency error of each helper.

According to \eqref{eq: freq_drift_limit}, the phase drift limits the slot duration from above. However, the slot duration cannot be too brief either. Since the SNR $\gamma_{2, i}$ is proportional to $T_{\rm s}$, it sets a minimum value for slot duration $T_{\rm s}$ required for reliable phase estimation so that 
\begin{equation}
     T_{\rm s} \geq \frac{\gamma_{2, {\rm min}}N_0}{P_{\rm h, min}},
     \label{eq:noise_limit}
\end{equation}
where $\gamma_{2, {\rm min}}$ is the minimum allowable input SNR for a desired performance level, while $P_{\rm h, min}$ indicates the minimum individual incident helper-tone power received by any A-mode node. In addition, because of the time it takes the signal from the A-mode helper to reach the tag, the phase offset that maximizes the helper node sum at the tag differs from the estimated one by $\theta_{\rm d} = 2 \pi \tau_k/T_{\rm s}$. Clearly, when $\tau_k \ll T_{\rm s}$ the phase delay $\theta_{\rm d}$ is negligible, which is why it was omitted earlier. Nevertheless, it contributes to the total phase difference and has to be accounted for when setting the SNR limit in \eqref{eq:noise_limit}.

Finally, combining \eqref{eq: freq_drift_limit} and \eqref{eq:noise_limit} we arrive at 
\begin{equation}
      \frac{\gamma_{2, {\rm min}}N_0}{P_{\rm h, max}}  \leq T_{\rm s} \leq \frac{\Delta \theta_{\rm max}}{(M-1)\omega_{{\rm er}, \rm max}},
      \label{eq:bounds_on_slot}
\end{equation}
which shows how the combination of noise, mobility and the number of helpers bounds the slot duration $T_{\rm s}$. In the following, we explore \eqref{eq:bounds_on_slot} in more detail.

%% file: figures/Kappa_PDF2.tex
\begin{tikzpicture}
\begin{axis}[width = 8cm, height = 5cm,
legend cell align = left,
grid = none,
enlarge x limits=false,
enlarge y limits=false,
ymin= 0,
ymax = 35,
xmin = 1.75, 
xmax = 2.01,
minor x tick num  = 1,
axis y line=left,
axis x line = bottom,
legend columns=2, 
legend style ={at={(0.5,-0.35)}, anchor= north, thin, font = \small,draw= gray},
 ylabel style={align=center}, ylabel={PDF $f_{\alpha_2}(\alpha_2)$}, xlabel = {Amplitude ratio $\alpha_2$}]
 
\addplot[thick, smooth, gray] coordinates {(0, 0)};
\addlegendentry{Theor. via \eqref{eq:final_a2} $\;$}

\addplot[thick, smooth, dashed, black] coordinates {(0, 0)};
\addlegendentry{Empir. via \eqref{eq:alpha}}

\addplot[very thick, smooth,  gray,  mark phase = 8, mark options = {scale = 0.9}, mblue] table[x index=0, y index=1, col sep=space] {data/PDF_alphai2gamma2_-5.txt};

\addplot[thick, smooth, black, dashed] table[x index=0, y index=2, col sep=space] {data/PDF_alpha2_gamma2_-5.txt};

\addplot[very thick, smooth,  gray,  mark phase = 7, mark options = {scale = 0.9}, mred] table[x index=0, y index=1, col sep=space] {data/PDF_alphai2gamma2_5.txt};

\addplot[thick, smooth, dashed, black] table[x index=0, y index=2, col sep=space] {data/PDF_alpha2_gamma2_5.txt};

\draw[-latex] (axis cs: 1.9, 25) node [left, text width=1.8cm, fill = white, font = \footnotesize] {$\gamma_2 = 5$~dB ($K_1 = 3.8$~dB)} to (axis cs: 1.98, 15) ;

\draw[-latex] (axis cs: 1.9, 12) node [left,text width=2cm, fill = white, font = \footnotesize] {$\gamma_2 = -5$~dB ($K_1 = -8.1$~dB)}  to (axis cs: 1.975, 4) ;

 \end{axis}
 \end{tikzpicture}

%% file: figures/PDF_alphan.tex
\begin{tikzpicture}
\begin{axis}[width = 8cm, height = 5cm,
legend cell align = left,
grid = none,
enlarge x limits=false,
enlarge y limits=false,
ymin= 0,
ymax = 9,
xmin = 0, 
xmax = 5.5,
minor x tick num  = 1,
axis y line=left,
axis x line = bottom,
cycle list name=linestyles*,
legend columns=2,
legend style ={at={(0.5,-0.35)}, anchor= north, font = \small,draw= gray},
 ylabel style={align=center}, ylabel={PDF $f_{\alpha_i}(\alpha_i)$}, xlabel = {Amplitude ratio $\alpha_i$}]

  \addplot+[smooth, black, solid] coordinates {(0, 1)};
 \addlegendentry{$\gamma_2 = 0$ dB $\;$}
 
  \addplot+[smooth, black, solid, very thick, dotted]
 table[x index=0, y index=1, col sep=space] {data/PDF_alphai2gamma2_-5.txt};
 \addlegendentry{$\gamma_2 = -5$ dB}
 
   \addplot+[smooth, mred, solid, thick] 
 table[x index=0, y index=1, col sep=space] {data/PDF_alphai2gamma2_0.txt};

\addplot+[smooth,  black, very thick,  solid,dotted] table[x index=0, y index=1, col sep=space] {data/PDF_alphai3gamma2_-5.txt};

\addplot+[smooth,  black, very thick,  solid,dotted] table[x index=0, y index=1, col sep=space] {data/PDF_alphai4gamma2_-5.txt};

  \addplot+[smooth,  black, very thick,  solid,dotted] table[x index=0, y index=1, col sep=space] {data/PDF_alphai5gamma2_-5.txt};
  
    \addplot+[smooth, thick] coordinates{(1, 0) (1, 0.1)};

\addplot+[smooth, mblue, solid, thick] table[x index=0, y index=1, col sep=space] {data/PDF_alphai3gamma2_0.txt};

\addplot+[smooth,  mpurple, solid, thick] table[x index=0, y index=1, col sep=space] {data/PDF_alphai4gamma2_0.txt};

  \addplot+[smooth, mgreen, solid, thick] table[x index=0, y index=1, col sep=space] {data/PDF_alphai5gamma2_0.txt};

\draw[-latex] (axis cs: 1.6, 8.5) node [left, fill = white,font = \footnotesize] {$i = 2$} to (axis cs: 1.95, 8.5) ; 

\draw[-latex] (axis cs: 3, 8.5) node [fill = white,font = \footnotesize] {$i = 3$} to (axis cs: 3, 6.4) ; 

\draw[-latex] (axis cs: 3.95, 7.5) node [fill = white,font = \footnotesize] {$i = 4$} to (axis cs: 3.95, 5.6) ; 

\draw[-latex] (axis cs: 4.95, 6.5) node [fill = white,font = \footnotesize] {$i = 5$} to (axis cs: 4.95, 5.5) ;

 \end{axis}
 \end{tikzpicture}

%% file: figures/REF_prcnt.tex
\begin{tikzpicture}
\begin{axis}[width = 8cm, height = 5cm,
legend cell align = left,
grid = both,
enlarge x limits=false,
enlarge y limits=false,
minor y tick num  = 1,
ymin = 1,
ymax = 2.5, 
axis y line*=left,
axis x line* = bottom,
legend style ={at={(0.67,0.53)}, anchor= north west, font = \small,draw= gray},
 ylabel style={align=center}, ylabel={ REF percentile $G_{\zeta_{\rm pa}}(p)$}, xlabel = {Number of helper nodes $M$}]

\addplot[thick, densely dotted,  mark = o, mark options = {solid}, mgreen] table[x index=0, y index=1, col sep=space] {data/REF_prcntl_gamma2-5.txt};
\addlegendentry{$p=10\%$}

\addplot[thick,  loosely dashdotted,  mark = x, mark options = {solid, scale= 1.5}, mpurple] table[x index=0, y index=2, col sep=space] {data/REF_prcntl_gamma2-5.txt};
\addlegendentry{$p=50\%$}

\addplot[thick,   black, dashed] table[x index=0, y index=4, col sep=space] {data/REF_prcntl_gamma2-5.txt};
\addlegendentry{$\zeta_{\rm coh}$}

\addplot[thick,  black] table[x index=0, y index=3, col sep=space] {data/REF_prcntl_gamma2-5.txt};
\addlegendentry{$\zeta_{\rm conv}$}

 \end{axis}

 \end{tikzpicture}

%% file: sec_Num.tex
\subsection{System Parameters}
In this section, we numerically evaluate proposed HR system with adaptive self-coherent auxiliary helper nodes and compare it against conventional HR without helpers. 
As a benchmark, we consider an X-band HR system reported in \cite{sto2020low} with operational parameters summarised in Table~\ref{tabl:params}. 
With an output transmit power of $P_{\rm r} = 10$Watt and Tx/Rx antenna gains of $15$dBi, its reported maximum detection range is 15m.

For modelling the behaviour of the harmonic tag, we consider a wire-based tag design from \cite{colpitts2004harmonic, lav2019design} where the tag consists of a single dipole antenna that  acts both as the receive antenna at $\omega_0$ and the transmit antenna at $ 2 \omega_0$, a Schottky diode and a parallel inductive loop. In our evaluation we use a Skyworks Schottky diode SMS7630-040 \cite{SkyWorksDiode} with parameters specified in Table~\ref{tabl:params_diode}. For simplicity, we also assume perfect matching conditions at both $\omega_0$ and $2 \omega_{0}$ such that 
$k_{\rm in} = k_{\rm out} = 1$.
To facilitate this, we assume that the tag antenna is a half-wavelength dipole at $\omega_0$ with an arm ratio of $(2:1)$, which, as outlined in \cite{aumann2012asymmetrical, law2020evaluation}, ensures a double resonant structure necessary to maximise the power transfer efficiency of the tag. Using an antenna-analysis tool MMANA-GAL \cite{MMANA-GAL}, we find that for a copper wire with a diameter of 0.31mm this yields $R_{\rm F} \approx 132\Omega$, $R_{\rm H} \approx 146\Omega$, $G_{\rm tag}(\omega_0) = 2.2$dBi, $G_{\rm tag}(2\omega_0) = 3.15$dBi. The tag antenna gains, together with the parameters from Table~\ref{tabl:params}, determine the uplink gain $h_{\rm u}(d)$ \eqref{eq:v_in} and the downlink gain $h_{\rm d}(d)$ \eqref{eq:received_sig}.

\begin{table}[t!]
%\vspace{-\baselineskip}
\renewcommand{\arraystretch}{1.3}
\caption{\small{Parameters of a HR system reported in \cite{sto2020low}}}
\label{table:summary}
\centering
\begin{tabular}{|l||c|c}
\hline
\bfseries Parameter &  \bfseries Value  \\
\hline  \hline
 Frequency  $f_0/2f_0$&  9.3/18.6 GHz \\
\hline
Receiver bandwidth $B_{\rm r}$ &  125 kHz\\
 \hline
 Output power $P_{\rm r}$ & 10 Watt \\
 \hline
 Tx/Rx antenna gain $G_{\rm tx}= G_{\rm rx}$  & 15 dBi\\
 \hline
 Rx noise figure $N_{\rm F}$ & 2.5 dB\\
\hline
\end{tabular}
\label{tabl:params}
\end{table}

\begin{table}[t!]
%\vspace{-\baselineskip}
\renewcommand{\arraystretch}{1.3}
\caption{\small{Parameters of a Schottky diode SMS7630-040 \cite{SkyWorksDiode}}}
\label{table:summary}
\centering
\begin{tabular}{|l||c|c}
\hline
\bfseries Parameter &  \bfseries Value  \\
\hline  \hline
Saturation current $I_{\rm s}$&  5 $\mu$A \\
\hline
    Ideality parameter $n_i$&  1.05 \\
 \hline
 Thermal voltage $V_{\rm T}$ & 26 mV \\
 \hline
 Coefficient $\rho$ & 0.024 \\
\hline
\end{tabular}
\label{tabl:params_diode}
\vspace*{-0.2cm}
\end{table}

Given an arbitrary transmit waveform $x(t)$, a transmit power $P_{\rm r}$ and a downlink gain $h_{\rm d}(d)$, we can determine the input voltage at the tag antenna $v_{\rm in}(t)$ according to \eqref{eq:v_in}. Throughout this paper, we made use of the quadratic small-signal approximation \eqref{eq:second_complx} to model the relation between the input voltage $v_{\rm in}(t)$ at the tag and the tag output current at the second harmonic $i_2(t)$. Here, we employ the more general explicit solution to \eqref{eq:diode_eq} instead \cite{lav2020two-region}:
\begin{equation}
    \tilde{i}_{\rm T}(t) = \left(\frac{ W_0\left(\rho e^{\left(\rho+\tilde{v}_{\rm in}(t)/n_{\rm i}V_{\rm T}\right)}\right)}{\rho} -1 \right)I_{\rm S},
    \label{eq:full_current}
\end{equation}
in which $W_0(\cdot)$ denotes the principal branch of the Lambert function \cite{corless1996lambertw}. Given \eqref{eq:full_current}, we compute $i_{2}(t)$ numerically as
\begin{equation}
     i_2(t) = a_2(t) -\jmath b_2(t),
     \label{eq:current_2}
\end{equation}
 where  $a_2(t), b_2(t)$ are the second harmonic Fourier coefficients of $\tilde{i}_{\rm T}(t)$ (see \cite{lav2020two-region} for details). As before, the output tag voltage is
 \begin{equation}
     v_{\rm out} = R_{\rm H} i_2(t).
 \end{equation}
Using \eqref{eq:full_current} allows us to test how well our quadratic model holds and evaluate how the proposed algorithm behaves outside the quadratic region.

\subsection{Conventional HR System}

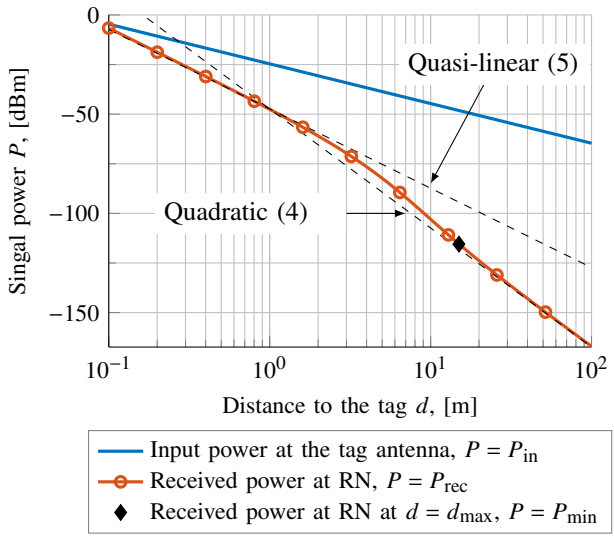
\begin{figure}[t!]
\vspace*{-\baselineskip}
    \centering
    \input{figures/PowVSDist.tex}
    \caption{\small{Signal power as a function of the distance to the tag $d$ in a conventional HR system without auxiliary helper nodes.}}
    \label{fig:PowVSDist}
    \vspace*{-0.2cm}
\end{figure}

We begin our study by determining operational conditions of a conventional X-band HR system without helper nodes with parameters specified in Table~\ref{tabl:params}. 
Fig.~\ref{fig:PowVSDist} shows input signal power at the tag antenna, $P_{\rm in} = h_{\rm u}^2(d)P_{\rm r}$, and at the harmonic radar receiver, $P_{\rm rec} = h_{\rm d}^2(d) R_{\rm H} I_2^2/2$, for a tone input, where $I_2$ denotes the amplitude of the second harmonic current. 
We can clearly observe the change in the curve slope of $P_{\rm rec}$ appearing around $d = 4$m. It illustrates the transition from the quasi-linear regime characteristic of the large-signal conditions to the quadratic regime in the small-signal region.
We also indicate here the point corresponding to the reported maximum detection range $d_{\rm max}= 15$m \cite{sto2020low}, which is clearly in the quadratic region. It determines the minimum received signal power of $P_{\rm min} = -115.5$dBm with a corresponding tag input power of $P_{\rm in} = -48$dBm. From this, we also compute a minimum required voltage amplitude at the tag, which is $A_{\rm r} = 2.43 n_i V_{\rm T} = 63$mV.
Finally, we determine the receiver noise power as $P_{\rm n} =  2 B_{\rm r} N_0  \approx -118$dBm, in which $N_0 = R_{\rm rx} k_{\rm B} T_{\rm n}(N_{\rm f}-1)$ where $T_{\rm n} = 290$K is the standard room temperature and $N_{\rm f}$ is the Rx noise figure in linear scale. %This results in the minimum required carrier to noise ratio of $2.5$dB. 

\begin{figure}[t!]
\vspace*{-\baselineskip}
\begin{subfigure}{0.49\textwidth}
%\vspace*{-0.2cm}
    \centering
   \input{figures/CDFalphaM_Ts.tex}
    \caption{\small{}}
    \label{fig:AlphaM_Prc_Ts}
\end{subfigure}
\begin{subfigure}{0.49\textwidth}
    \centering
   \input{figures/CDFalphaM_d.tex}
        \vspace*{-0.4cm}
    \caption{\small{}}
    \label{fig:AlphaM_Prc_d}
\end{subfigure}
 \caption{\small{Normalized amplitude ratio percentile $G_{\tilde{\alpha}_M} (p)$ as a function of the  input SNR $\gamma_2$ (a) at a fixed distance $d=d_{\rm max}$  and (b) at a fixed slot duration $T_{\rm s} = 1$ns.}}
     \label{fig:AlphaM_Prc}
      \vspace*{-0.3cm}
\end{figure}
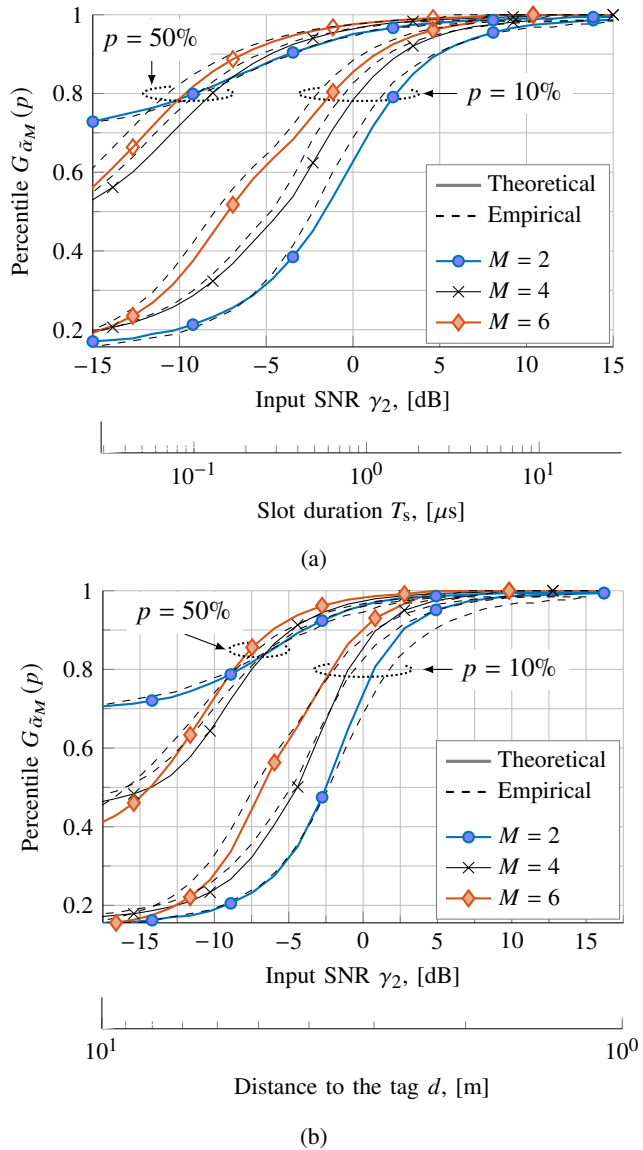

\begin{figure*}[t!]
\vspace*{-\baselineskip}
\begin{subfigure}{0.49\textwidth}
    \centering
    \centering
   \input{figures/SlotDurationvsAlphaM.tex}
 \caption{}
     \label{fig:Ts_AlphaM_Prc}
     \end{subfigure}
 \begin{subfigure}{0.49\textwidth}
    \centering
    \centering
   \input{figures/SlotDurationvsAlphaM_freq.tex}
 \caption{}
     \label{fig:Ts_AlphaM_Prc_freq}
     \end{subfigure}
     \caption{\small{Slot duration required for 10th percentile of helper-based HR amplitude $G_{2\alpha_M} (10)$ to exceed conventional HR amplitude $M+1$: (a) minimum required $T_{\rm s}$  in the presence of the propagation delay $\theta_{\rm d}=2 \pi d_{\rm max} /cT_{\rm s}$ and (b) maximum allowed $T_{\rm s}$  in the presence of the frequency error $\omega_{\rm err} = 10^{-6} p_{\rm e} \omega_0$.}}
        \label{eq:slotduration}
                \vspace*{-0.3cm}
\end{figure*}
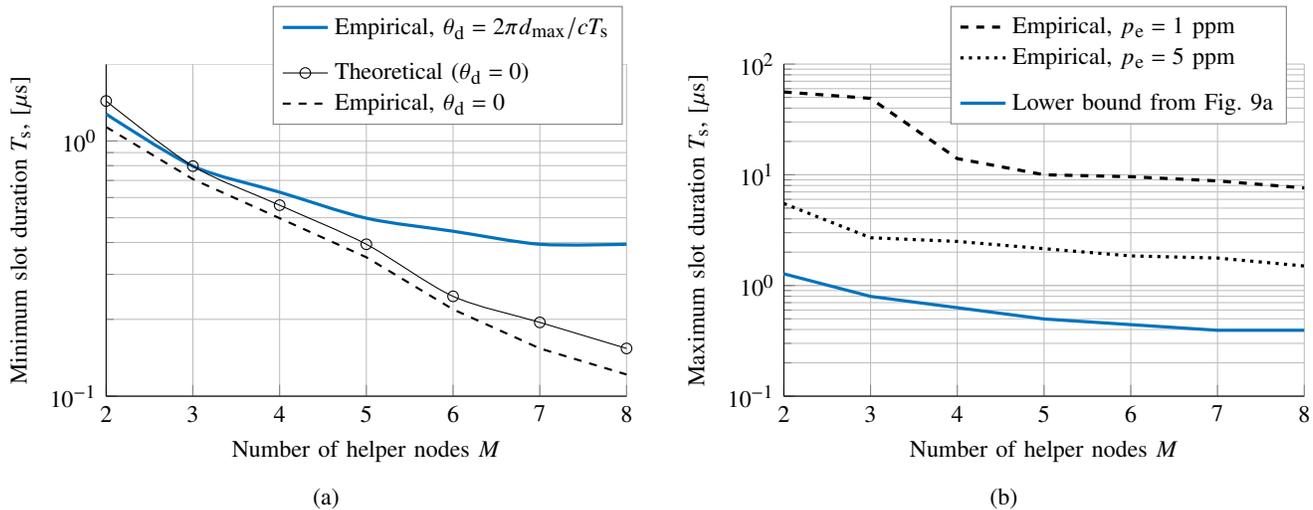

\subsection{HR System with Auxiliary Helper Nodes}
\label{sec:num3}

This section examines the benefits of introducing helpers into the operational scenario described above. As before, we assume for simplicity that all helper tones arrive at the tag with the same amplitude $A_{\rm r}$ so that when they are perfectly phase-aligned at the tag the composite helper node amplitude $A_{\rm h}$ is equal to $MA_{\rm r}$ and the amplitude ratio at the end of the phase adjustment interval is $\alpha_M = M$. First, we present the statistical distributions of the normalized amplitude ratio $\tilde{\alpha}_M = \alpha_M/M \in [0, 1]$. Previously, we saw that $f_{\alpha_M}(\alpha_M)$ depends on the input SNR\footnote{In light of using \eqref{eq:full_current} for determining the tag output, we define $\gamma_2$ here as the SNR at the output of the third Fourier integrator \eqref{eq:Fintegral3} during the first phase adjustment slot,  i.e., $\gamma_2 = \gamma_{2, 1} = |\dot{G}_{2, 1}|^2/T_{\rm s}N_0$.} $\gamma_2$, which in turn depends both on the phase adjustment slot duration $T_{\rm s}$ via the integration time in \eqref{eq:Fintegral3} and the distance to the tag $d_{\rm r}$ via the amplitude of the receiver input. Suppose now that the tag is positioned at the maximum range of the conventional system such that $d_{\rm r} = d_{\rm max}$. Then, the amplitude $A_{\rm r}$ is fixed and $\gamma_2$ is determined by the slot duration $T_{\rm s}$.  Fig.~\ref{fig:AlphaM_Prc_Ts} shows $10$th and $50$th percentiles $G_{\tilde{\alpha}_M} (p)$ of the normalized amplitude ratio $\tilde{\alpha}_M$ as a function of the input SNR $\gamma_2$, together with the values of $T_{\rm s}$ corresponding to the SNRs. We observe good correspondence between the empirical results obtained using \eqref{eq:full_current} and the theoretical ones derived from the asymptotic quadratic model \eqref{eq:second_complx}. Fig.~\ref{fig:AlphaM_Prc_Ts} also demonstrates that the phase adjustment performance improves both with $M$ and the input SNR. For instance, when $M=2$ and the slot duration is greater than one microsecond (i.e., $\gamma_2>0$), $90\%$ of the outcomes achieve more than $60\%$ of the maximum possible amplitude amplification (i.e., $\alpha_M$ is greater than $0.6M=1.2$) while for $M=4$ it exceeds $85\%$ (i.e, $\alpha_M$ is greater than $0.85M=3.4$). To test how well our approach performs outside of the quadratic region, we fix the slot duration $T_{\rm s}$ and change the distance to the tag $d$ instead. Note that the slot duration is set to $T_{\rm s} = 1$ns, an extremely low value, solely to avoid excessively large SNR values when the distance is small and the tag operates  far above the quadratic region.   Fig.~\ref{fig:AlphaM_Prc_d} demonstrates the results. It indicates that the move towards the quasi-linear ($d<<d_{\rm max}$) region does not significantly affect the ability of our phase adaptation algorithm to phase-align helper signals at the tag.

Next, we return to Section~\ref{sec:SLD} considerations and determine the upper and lower bounds on slot duration that allow the helper-based HR system to outperform conventional HR, when both systems are given the same total transmit power budget. To this end, we again consider system performance at the maximum range by setting the tag distance to $d_{\rm max}$, as it is our primary region of interest. First, we study the lower bound \eqref{eq:noise_limit} on slot duration $T_{\rm s}$. To account for the propagation delay in the sweep function $\varphi(t)$, we introduce an additional error $\theta_{\rm d} = 2 \pi d_{\rm max} /cT_{\rm s}$ into the phase offset estimate \eqref{eq:phase_off_est}. As our performance metric we choose the amplitude of the helper-based system measured by the pessimistic 10th percentile point, i.e., $G_{2{\alpha}_M}(10)$.  
Recall that the amplitude of the signal into the tag is proportional to $2\alpha_M$ for helper-assisted HR \eqref{eq:ranging_pwr} and to $M+1$ for conventional HR \eqref{eq:convent_power}.
 Thus, Fig.~\ref{fig:AlphaM_Prc_Ts} shows the minimum slot duration required for that 10th percentile amplitude to exceed the conventional HR amplitude, i.e., $G_{2\alpha_M}(10) > M+1$. 
For comparison, we also show theoretical and empirical curves obtained under the initial assumption that ${\theta}_{\rm d} = 0$. 
Fig.~\ref{fig:AlphaM_Prc_Ts} suggests that a slot duration of about $1\mu$s ensures that the SNR boost of the helper-node system exceeds that of a comparable conventional system in $90\%$ of the cases. The slot duration can be further reduced by accepting a lower percentage of  outcomes in which $2\alpha_M > M+1$, e.g., by considering a $50$th percentile instead.

We now turn to evaluating the upper bound \eqref{eq: freq_drift_limit} on slot duration $T_{\rm s}$ that still allows the pessimistic 10th amplitude percentile in helper-based HR to exceed the amplitude of the conventional HR. We introduce a frequency error $\omega_{\rm er}$ as a portion of the fundamental carrier frequency $\omega_0$, so that $\omega_{\rm er}/\omega_0 = 10^{-6} p_{\rm e} $ where $p_{\rm e}$ is the frequency instability expressed in parts per million (ppm). Given $\omega_{\rm er}$, the maximum phase error between any two helpers over the slot duration is $2\omega_{\rm er}T_{\rm s}$. Fig.~\ref{fig:Ts_AlphaM_Prc_freq} shows the maximum slot duration $T_{\rm s}$ that permits $G_{2\alpha_M}(10\%) > M+1$ as a function of the number of helpers $M$, for two values of $p_{\rm e}$. As expected, the upper bound begins to approach the lower one from Fig.~\ref{fig:Ts_AlphaM_Prc} when the frequency error is increased. As a result, for large enough frequency errors there might be no suitable slot duration that fulfills \eqref{eq:bounds_on_slot} for the performance metric of choice, so one would have to accept a higher probability of low-power outcomes. On the other hand, Fig.~\ref{eq:slotduration} shows that when the frequency error is within several ppm, the interplay between noise, frequency instability and the number of helpers provides enough room to choose $T_{\rm s}$ that ensures that the probability of low-power outcome does not exceed a desired value ($10\%$ in our case). 

\begin{figure}[t!]
%\vspace{-\baselineskip}
   \centering
  \input{figures/REF_FINAL.tex}
  \vspace*{0.2cm}
    \caption{\small{CDFs of the REF $\zeta_{\rm pa}$ of the phase-adaptive helper-based system with the slot duration of $1\mu$s and a frequency instability of 1 ppm. Individual circle and rhomboid markers show the corresponding REFs of a fully coherent helper-based system $\zeta_{\rm coh} = \sqrt[3]{2M}$ and a  brute-force conventional system without helpers $\zeta_{\rm conv} = \sqrt[3]{M+1}$, respectively.}}
   \label{fig:REF_final}
    \vspace*{-0.3cm}
\end{figure}
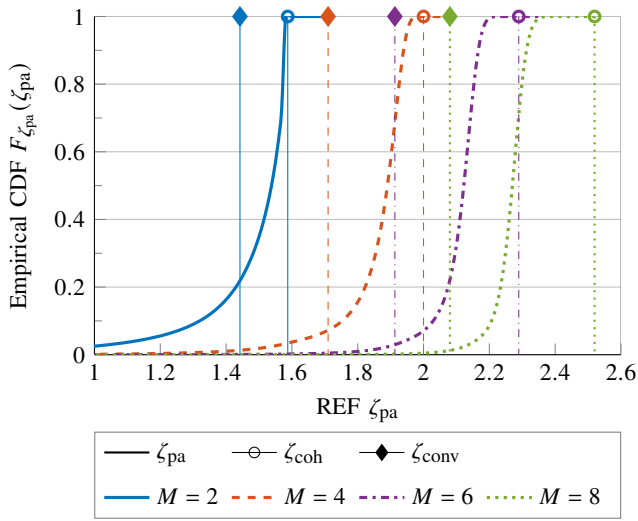

Our third, and final, assessment is to fix the slot duration and evaluate the REF  of our  phase-adaptive helper-based system ($\zeta_{\rm pa}= \sqrt[3]{2\alpha_M}$) taking into account both the phase delay $\theta_{\rm d}$ and the frequency error $\omega_{\rm err}$. Fig.~\ref{fig:REF_final} shows the CDF of  $\zeta_{\rm pa}$ for $T_{\rm s} = 1\mu$s (i.e., $\gamma_2 = 0.4$dB at $d_{\rm max}$) and $p_{\rm e} = 1$ ppm. For comparison, it also provides corresponding values for the fully coherent system ($\zeta_{\rm coh} = \sqrt[3]{2M}$, marked by circles) and the brute-force conventional system ($\zeta_{\rm conv} = \sqrt[3]{M+1}$, marked by filled rhomboids). We observe that setting the slot duration within the limits determined by the performance bounds from Fig.~\ref{eq:slotduration} results in the phase-adaptive system outperforming the conventional HR with equivalent total transmit power with high probability, i.e., in more than $80\%$ of the cases. Furthermore, the performance gap between the two increases with $M$. These results show that helpers can be added incrementally in order to achieve the desired range or SNR increase, as long as the system design satisfies \eqref{eq:bounds_on_slot}.

%% file: figures/PowVSDist.tex
\begin{tikzpicture}
\begin{axis}[width = 8cm, height = 6cm,
legend cell align = left,
grid = both,
xmode = log,
xmax = 100,
ymax = 0,
xmin = 0.1,
minor y tick num  = 1,
legend style ={at={(0.5,-0.25)},anchor=north,draw = gray, font=\small, thin}, 
font = \normalsize,
enlarge x limits=false,
enlarge y limits=false,
axis y line*=left,
axis x line* = bottom,
 ylabel style={align=center}, ylabel={Singal power $P$, [dBm]}, xlabel = {Distance to the tag $d$, [m]}]

\addplot+[very thick, smooth, mark repeat = 10, mark = none, mblue] table[x index=0, y index=1, col sep=space] {data/Mag_vs_dist.txt};
\addlegendentry{Input power at the tag antenna, $P = P_{\rm in}$}

\addplot+[very thick, smooth, mark repeat = 20, mark = o, mark options = {solid}, mred] table[x index=0, y index=2, col sep=space] {data/Mag_vs_dist.txt};
\addlegendentry{Received power at RN, $P =P_{\rm rec}$}

\addplot+[black, only marks, mark = diamond*, mark options = {scale = 1.5}] coordinates {(15, -115.5)};
\addlegendentry{Received power at RN at $d = d_{\rm max}$, $P = P_{\rm min}$}

\addplot+[smooth, mark repeat = 20, mark = none, dashed, black, forget plot] table[x index=0, y index=3, col sep=space] {data/Mag_vs_dist.txt};

\addplot+[smooth, mark repeat = 20, mark = none, dashed, black, forget plot] table[x index=0, y index=4, col sep=space] {data/Mag_vs_dist.txt};

\draw[-latex] (axis cs: 3, -100) node [left, text width=2.3cm, fill = white] {Quadratic \eqref{eq:second_complx} } to (axis cs: 7, -100) ;

\draw[-latex] (axis cs: 25, -25) node [ text width=2.3cm, fill = white] {Quasi-linear \eqref{eq:current_second_large} } to (axis cs: 10, -85) ;

 \end{axis}
 \end{tikzpicture}
 
 %\end{center}

%% file: figures/CDFalphaM_Ts.tex
\tikzset{
    partial ellipse/.style args={#1:#2:#3}{
        insert path={+ (#1:#3) arc (#1:#2:#3)}
    }
}

\begin{tikzpicture}
\begin{axis}[width = 8.5cm, height = 6cm,
legend cell align = left,
grid = both,
enlarge x limits=false,
enlarge y limits=false,
minor y tick num  = 1,
axis y line*=left,
axis x line* = bottom,
xmin = -15,
legend style ={at={(0.64,0.55)}, anchor= north west, font = \small,draw= gray},
 ylabel style={align=center}, ylabel={Percentile $G_{\tilde{\alpha}_M}(p)$}, xlabel = {Input SNR $\gamma_2$, [dB]}]

\addplot[very thick, smooth, gray] coordinates {(0, 1)};
\addlegendentry{Theoretical}

\addplot[thick, smooth, dashed, black] coordinates {(0, 1)};
\addlegendentry{Empirical}

\addplot[thick, smooth, dashed, white] coordinates {(0, 1)};
\addlegendentry{$\;$}

\addplot[thick,  mark = *, mark options = {solid, fill = blue!50!white}, mark repeat = 5, mblue] table[x index=0, y index=1, col sep=space] {data/AlphaM_prcnt.txt};
\addlegendentry{$M=2$}

\addplot[thin,   black, mark = x, mark options = {solid, scale = 1.5, fill = gray}, mark repeat = 5, mark phase = 2] table[x index=0, y index=9, col sep=space] {data/AlphaM_prcnt.txt};
\addlegendentry{$M=4$}

\addplot[thick,   mred, mark = diamond*, mark options = {solid, scale = 1.5, fill = mred!50!white}, mark repeat = 5, mark phase = 3] table[x index=0, y index=17, col sep=space] {data/AlphaM_prcnt.txt};
\addlegendentry{$M=6$}

\addplot[ thin,  dashed, black] table[x index=0, y index=2, col sep=space] {data/AlphaM_prcnt.txt};
\addplot[thin,  dashed, black] table[x index=0, y index=10, col sep=space] {data/AlphaM_prcnt.txt};
\addplot[ thin,  dashed, black] table[x index=0, y index=18, col sep=space] {data/AlphaM_prcnt.txt};

\addplot[thick,  mark = *, mark options = {solid, fill = blue!50!white}, mark repeat = 5, mblue] table[x index=0, y index=3, col sep=space] {data/AlphaM_prcnt.txt};
\addplot[thin,   black, mark = x, mark options = {solid, scale = 1.5}, mark repeat = 5, mark phase = 2] table[x index=0, y index=11, col sep=space] {data/AlphaM_prcnt.txt};

\addplot[thick,   mred, mark = diamond, mark options = {solid, scale = 1.5, fill = mred!50!white}, mark repeat = 5, mark phase = 3] table[x index=0, y index=19, col sep=space] {data/AlphaM_prcnt.txt};

\addplot[thin,  dashed, black] table[x index=0, y index=4, col sep=space] {data/AlphaM_prcnt.txt};
\addplot[thin,  dashed, black] table[x index=0, y index=12, col sep=space] {data/AlphaM_prcnt.txt};
\addplot[thin,  dashed, black] table[x index=0, y index=20, col sep=space] {data/AlphaM_prcnt.txt};

\draw[thick, densely dotted] (axis cs: 0.35, 0.8) [partial ellipse=30:-230:0.8cm and 0.1cm];
\draw[-latex] (axis cs: 6, 0.8) node [right,text width=1.3cm, fill = white]{$p=10\%$}  to (axis cs: 4, 0.8) ;

\draw[thick, densely dotted] (axis cs: -9.5, 0.8) [partial ellipse=30:-250:0.6cm and 0.1cm];

\draw[-latex] (axis cs: -11.6, 0.99) node [below,text width=1.3cm, fill = white]{$p=50\%$}  to (axis cs:-11.6, 0.82) ;

 \end{axis}
 \end{tikzpicture}
 \begin{tikzpicture}
 \begin{axis}[height=2cm, width = 8.5cm, 
grid = none,
xmode = log,
xmax = 30,
enlarge x limits=false,
ymajorticks=false,
 axis y line=left,
 y axis line style=-,
 axis x line* = bottom,
xticklabel style={below}, very thin,
xlabel={Slot duration $T_{\rm s}$, [$\mu$s]}, 
legend style ={at={(-0.185,0)}, anchor= west, font = \small, fill = none, draw=none},
 font  = \small]

\addplot+[white, mark = none, mark options = {mark repeat = 3, mark phase =2}] table[x index=1, y index=0, col sep=space] {data/Gamma2_Ts.txt};
\addlegendentry{$\,$}

\end{axis}

 \end{tikzpicture}

%% file: figures/CDFalphaM_d.tex
\tikzset{
    partial ellipse/.style args={#1:#2:#3}{
        insert path={+ (#1:#3) arc (#1:#2:#3)}
    }
}

\begin{tikzpicture}
\begin{axis}[width = 8.5cm, height = 6cm,
legend cell align = left,
grid = both,
enlarge x limits=false,
enlarge y limits=false,
minor y tick num  = 1,
minor x tick num  = 1,
axis y line*=left,
xmax =17.5,
xmin= -17.5,
axis x line* = bottom,
legend style ={at={(0.64,0.55)}, anchor= north west, font = \small,draw= gray},
 ylabel style={align=center}, ylabel={Percentile $G_{\tilde{\alpha}_M}(p)$}, xlabel = {Input SNR $\gamma_2$, [dB]}]

\addplot[very thick, smooth, gray] coordinates {(0, 1)};
\addlegendentry{Theoretical}

\addplot[thick, smooth, dashed, black] coordinates {(0, 1)};
\addlegendentry{Empirical}

\addplot[thick, smooth, dashed, white] coordinates {(0, 1)};
\addlegendentry{$\;$}

\addplot[thick,  mark = *, mark options = {solid, fill = blue!50!white}, mark repeat = 4, mblue] table[x index=0, y index=1, col sep=space] {data/AlphaM_prcnt_d.txt};
\addlegendentry{$M=2$}

\addplot[ thin,   black, mark = x, mark options = {solid, scale = 1.5}, mark repeat = 4, mark phase = 2] table[x index=0, y index=9, col sep=space] {data/AlphaM_prcnt_d.txt};
\addlegendentry{$M=4$}

\addplot[thick,   mred, mark = diamond*, mark options = {solid, scale = 1.5, fill = mred!50!white}, mark repeat = 4, mark phase = 3] table[x index=0, y index=17, col sep=space] {data/AlphaM_prcnt_d.txt};
\addlegendentry{$M=6$}

\addplot[thin,  dashed, black] table[x index=0, y index=2, col sep=space] {data/AlphaM_prcnt.txt};
\addplot[ thin,  dashed, black] table[x index=0, y index=10, col sep=space] {data/AlphaM_prcnt_d.txt};
\addplot[ thin,  dashed, black] table[x index=0, y index=18, col sep=space] {data/AlphaM_prcnt_d.txt};

\addplot[thick,  mark = *, mark options = {solid, fill = blue!50!white}, mark repeat = 4, mblue] table[x index=0, y index=3, col sep=space] {data/AlphaM_prcnt_d.txt};

\addplot[ thin,   black, mark = x, mark options = {solid, scale = 1.5}, mark repeat = 4, mark phase = 2] table[x index=0, y index=11, col sep=space] {data/AlphaM_prcnt_d.txt};

\addplot[thick,   mred, mark = diamond*, mark options = {solid, scale = 1.5, fill = mred!50!white}, mark repeat = 3, mark phase = 3] table[x index=0, y index=19, col sep=space] {data/AlphaM_prcnt_d.txt};

\addplot[thin,  dashed, black] table[x index=0, y index=4, col sep=space] {data/AlphaM_prcnt.txt};
\addplot[ thin,  dashed, black] table[x index=0, y index=12, col sep=space] {data/AlphaM_prcnt_d.txt};
\addplot[ thin,  dashed, black] table[x index=0, y index=20, col sep=space] {data/AlphaM_prcnt_d.txt};

\draw[thick, densely dotted] (axis cs: 0, 0.8) [partial ellipse=30:-230:0.65cm and 0.1cm];
\draw[-latex] (axis cs: 6, 0.8) node [right,text width=1.3cm, fill = white]{$p=10\%$}  to (axis cs: 4, 0.8) ;

\draw[thick, densely dotted] (axis cs: -7, 0.85) [partial ellipse=30:-250:0.4cm and 0.1cm];

\draw[-latex] (axis cs: -12, 0.89) node [above,text width=1.3cm, fill = white]{$p=50\%$}  to (axis cs:-9.2, 0.85) ;

 \end{axis}
 \end{tikzpicture}
 \begin{tikzpicture}
 \begin{axis}[height=2cm, width = 8.5cm, 
grid = none,
xmode = log,
ymajorticks=false,
 axis y line=left,
 y axis line style=-,
 x dir=reverse,
 axis x line* = bottom,
 xmin =1, 
 xmax = 10,
 xtick = {1, 10},
xticklabel style={below}, very thin,
xlabel={Distance to the tag $d$, [m]}, 
legend style ={at={(-0.22,0)}, anchor= west, font = \small, fill = none, draw=none},
 font  = \small]

\addplot+[white, mark = none, mark options = {mark repeat = 3, mark phase =2}] table[x index=0, y index=1, col sep=space] {data/Gamma2_d.txt};
\addlegendentry{$\;$}

\end{axis}

 \end{tikzpicture}

%% file: figures/SlotDurationvsAlphaM.tex
\begin{tikzpicture}
\begin{axis}[width = 8.5cm, height = 6cm,
legend cell align = left,
grid = both,
ymode = log,
enlarge x limits=false,
enlarge y limits=false,
minor y tick num  = 1,
axis y line*=left,
xmin = 2,
ymin = 0.1,
ymax =2,
axis x line* = bottom,
legend style ={at={(0.32,1.175)}, anchor= north west, font = \small,draw= gray},
 ylabel style={align=center}, xlabel={Number of helper nodes $M$}, ylabel = {Minimum slot duration $T_{\rm s}$, [$\mu$s]}]

\addplot[smooth, very thick,  solid, mark options = {solid}, mblue] table[x index=0, y index=3, col sep=space] {data/Ts_M_AlphaM_prcnt.txt};
\addlegendentry{Empirical, $\theta_{\rm d} = 2 \pi d_{\rm max} /cT_{\rm s}$}

\draw[lightgray] (axis cs: 2, 2) -- (axis cs: 8, 2);

\addplot+[thick, smooth, white, solid, mark = none, mark options = {scale = 1.5, solid}] coordinates {(1, 1)};
\addlegendentry{$\;$}

\addplot[smooth, thin,  mark = o, mark options = {solid}, black] table[x index=0, y index=1, col sep=space] {data/Ts_M_AlphaM_prcnt.txt};
\addlegendentry{Theoretical ($\theta_{\rm d} = 0$)}

\addplot[thick, dashed, mark options = {solid}, black] table[x index=0, y index=5, col sep=space] {data/Ts_M_AlphaM_prcnt.txt};
\addlegendentry{Empirical, $\theta_{\rm d} = 0$}

 \end{axis}
 \end{tikzpicture}

%% file: figures/SlotDurationvsAlphaM_freq.tex
\begin{tikzpicture}
\begin{axis}[width = 8.5cm, height = 6cm,
legend cell align = left,
grid = both,
ymode = log,
enlarge x limits=false,
enlarge y limits=false,
minor y tick num  = 1,
axis y line*=left,
xmin = 2,
ymin = 0.1,
ymax = 100,
axis x line* = bottom,
legend style ={at={(0.32,1.175)}, anchor= north west, font = \small,draw= gray},
 ylabel style={align=center}, xlabel={Number of helper nodes $M$}, ylabel = {Maximum slot duration $T_{\rm s}$, [$\mu$s]}]

\addplot[very thick,  dashed, mark options = {solid}, black] table[x index=0, y index=1, col sep=space] {data/Ts_M_AlphaM_1ppm_10prcnt.txt};
\addlegendentry{Empirical, $p_{\rm e}=1$ ppm}

\addplot[very thick, dotted, mark options = {solid}, black] table[x index=0, y index=1, col sep=space] {data/Ts_M_AlphaM_5ppm_10prcnt.txt};
\addlegendentry{Empirical, $p_{\rm e}=5$ ppm}

\addplot+[thick, smooth, white, solid, mark = none, mark options = {scale = 1.5, solid}] coordinates {(1, 1)};
\addlegendentry{$\;$}

\addplot[very thick,  solid, mark options = {solid}, mblue] table[x index=0, y index=3, col sep=space] {data/Ts_M_AlphaM_prcnt.txt};

\addlegendentry{Lower bound from Fig.~\ref{fig:Ts_AlphaM_Prc}}

 \end{axis}
 \end{tikzpicture}

%% file: figures/REF_FINAL.tex
\begin{tikzpicture}
% let both axes use the same layers
\pgfplotsset{set layers}
\begin{axis}[width = 7cm, height = 4.5cm,
legend cell align = left,
ymajorgrids=true,
scale only axis,
xmin = 1,
xmax = 2.6,
minor y tick num  = 1,
axis y line*=left, 
legend style ={at={(0.5,-0.485)}, anchor= south, draw =gray}, legend columns = 4,
font = \normalsize,
enlarge x limits=false,
enlarge y limits=false,
axis x line* = bottom,
 ylabel style={align=center}, ylabel={Empirical CDF $F_{\zeta_{\rm pa}}(\zeta_{\rm pa})$}, xlabel = {REF $\zeta_{\rm pa}$}
]

\addplot+[thick, smooth, black, solid, mark = none] coordinates {(0, 0)};
\addlegendentry{$\zeta_{\rm pa}$}

\addplot+[thin, smooth, black, solid, mark = o] coordinates {(0, 0)};
\addlegendentry{$\zeta_{\rm coh}$}

\addplot+[thin, smooth, black, solid, mark = diamond*, mark options = {scale = 1.5, solid}] coordinates {(0, 0)};
\addlegendentry{$\zeta_{\rm conv}$}

\addplot+[thick, smooth, white, solid, mark = none, mark options = {scale = 1.5, solid}] coordinates {(0, 0)};
\addlegendentry{$\;$}

\addplot+[thick, smooth, white, solid, mark = none, mark options = {scale = 1.5, solid}] coordinates {(0, 0)};
\addlegendentry{$\;$}
\addplot+[thick, smooth, white, solid, mark = none, mark options = {scale = 1.5, solid}] coordinates {(0, 0)};
\addlegendentry{$\;$}
\addplot+[thick, smooth, white, solid, mark = none, mark options = {scale = 1.5, solid}] coordinates {(0, 0)};
\addlegendentry{$\;$}
\addplot+[thick, smooth, white, solid, mark = none, mark options = {scale = 1.5, solid}] coordinates {(0, 0)};
\addlegendentry{$\;$}

\addplot+[very thick, smooth, solid, mblue, mark = none] table[x index=0, y index=1, col sep=space] {data/CDF_REF_M2.txt};
\addlegendentry{$M=2 \;$}

\addplot+[very thick, dashed, smooth, mred, mark = none] table[x index=0, y index=1, col sep=space] {data/CDF_REF_M4.txt};
\addlegendentry{$M=4 \;$}

\addplot+[very thick, dashdotted, smooth, mpurple, mark = none] table[x index=0, y index=1, col sep=space] {data/CDF_REF_M6.txt};
\addlegendentry{$M=6 \;$}

\addplot+[very thick, dotted, smooth, mgreen, mark = none] table[x index=0, y index=1, col sep=space] {data/CDF_REF_M8.txt};
\addlegendentry{$M=8 \;$}

\addplot+[very thick, smooth, mblue, mark = o] coordinates {(1.5874, 1)};
\addplot+[very thick, smooth, mred, mark = o] coordinates {(2, 1)};
\addplot+[very thick, smooth, mpurple, mark = o] coordinates {(2.2894, 1)};
\addplot+[very thick, smooth, mgreen, mark = o] coordinates {(2.5198, 1)};

\addplot+[thin, smooth, solid, mblue, mark = none] coordinates {(1.5874, 1) (1.5874, 0)};
\addplot+[thin, smooth,  mred,dashed, mark = none] coordinates {(2, 1) (2, 0)};
\addplot+[thin, smooth,  mpurple,dashdotted,  mark = none] coordinates {(2.2894, 1) (2.2894, 0)};
\addplot+[thick, smooth, mgreen,  dotted, mark = none] coordinates {(2.5198, 1) (2.5198, 0)};

\addplot+[very thick, smooth, mblue, mark = diamond*, mark options = {scale = 1.5, solid}] coordinates {(1.4422, 1)};
\addplot+[very thick, smooth, mred, mark = diamond*, mark options = {scale = 1.5, solid}] coordinates {(1.71, 1)};
\addplot+[very thick, dashdotted, mpurple, mark =diamond*, mark options = {scale = 1.5, solid}] coordinates {(1.9129, 1)};
\addplot+[very thick, dotted, mgreen, mark =diamond*, mark options = {scale = 1.5, solid}] coordinates {(2.0801, 1)};

\addplot+[thin, smooth, solid, mblue, mark = none] coordinates {(1.4422, 1) (1.4422, 0)};
\addplot+[thin, smooth, dashed, mred, mark = none] coordinates {(1.71, 1) (1.71, 0)};
\addplot+[thin, smooth, mpurple, dashdotted, mark = none] coordinates {(1.9129, 1) (1.9129, 0)};
\addplot+[thick, smooth, mgreen, dotted, mark = none] coordinates {(2.0801, 1) (2.0801, 0)};

\addplot+[thin, smooth, gray, solid, mark = none] coordinates {(2.6, 1) (2.6, 0)};

\end{axis}
\end{tikzpicture}